\documentclass[10pt]{article}

\usepackage[letterpaper,margin=1.03in,footskip=0.55in]{geometry}
\usepackage[sc]{mathpazo}
\usepackage[scaled=0.90]{helvet}
\usepackage{courier}
\usepackage[T1]{fontenc}
\usepackage[utf8]{inputenc}
\usepackage{microtype}
\usepackage{amsmath,amssymb,amsthm}
\usepackage{booktabs}
\usepackage{listings}
\usepackage{xcolor}
\usepackage{enumitem}
\usepackage{graphicx}
\usepackage{float}
\usepackage[font=small,labelfont=bf,labelsep=period,skip=6pt]{caption}
\usepackage{titlesec}
\usepackage{fancyhdr}
\usepackage[round]{natbib}

\usepackage{tikz}
\usetikzlibrary{arrows.meta,positioning,calc,fit,backgrounds}
\usepackage[most]{tcolorbox}
\PassOptionsToPackage{hyphens}{url}
\usepackage[colorlinks=true,allcolors=linkink]{hyperref}
\definecolor{ink}{RGB}{30,33,38}
\definecolor{linkink}{RGB}{42,72,120}
\definecolor{slate}{RGB}{88,101,120}
\definecolor{fog}{RGB}{176,186,199}
\definecolor{cardgray}{RGB}{248,249,251}
\definecolor{cardteal}{RGB}{237,246,244}
\definecolor{cardamber}{RGB}{250,245,235}
\definecolor{teal}{RGB}{17,104,96}
\definecolor{amber}{RGB}{158,88,22}
\definecolor{rowa}{RGB}{120,144,156}
\definecolor{rowb}{RGB}{17,104,96}
\definecolor{rowc}{RGB}{100,116,139}
\definecolor{rowd}{RGB}{109,76,140}

\lstdefinestyle{benchmarkprompt}{
  basicstyle=\ttfamily\scriptsize,
  backgroundcolor=\color{cardgray},
  frame=single,
  rulecolor=\color{fog},
  breaklines=true,
  breakatwhitespace=false,
  columns=fullflexible,
  keepspaces=true,
  showstringspaces=false,
  xleftmargin=3pt,
  xrightmargin=3pt,
  aboveskip=7pt,
  belowskip=9pt
}

\titleformat{\section}{\Large\bfseries}{\thesection}{0.7em}{}
\titleformat{\subsection}{\large\bfseries}{\thesubsection}{0.6em}{}
\titleformat{\subsubsection}{\normalsize\bfseries}{\thesubsubsection}{0.6em}{}
\titleformat{\paragraph}[runin]{\bfseries}{}{0em}{}
\titlespacing*{\section}{0pt}{16pt plus 3pt minus 2pt}{7pt}
\titlespacing*{\subsection}{0pt}{11pt plus 2pt minus 2pt}{5pt}
\titlespacing*{\subsubsection}{0pt}{9pt plus 2pt minus 1pt}{4pt}
\titlespacing*{\paragraph}{0pt}{7pt plus 2pt}{1.1em}

\fancypagestyle{plain}{\fancyhf{}\fancyfoot[C]{\footnotesize\thepage}}

\newtheoremstyle{defsty}{7pt}{7pt}{\itshape}{}{\bfseries}{.}{0.6em}{}
\theoremstyle{defsty}
\newtheorem{definition}{Definition}

\definecolor{thesisbg}{RGB}{238,243,250}
\definecolor{thesisframe}{RGB}{203,216,235}
\newcommand{\thesisbox}[1]{%
  \begin{tcolorbox}[enhanced, colback=thesisbg, colframe=thesisframe,
    boxrule=0.5pt, arc=2.5pt, left=8pt, right=8pt, top=5pt, bottom=5pt,
    before skip=9pt, after skip=9pt]\itshape #1\end{tcolorbox}}
\newcommand{\term}[1]{\textsc{#1}}
\newcommand{\qtext}[1]{``\textit{#1}''}
\tcolorboxenvironment{definition}{enhanced, breakable,
  colback=cardteal, colframe=teal!38, boxrule=0.5pt, arc=2pt,
  left=6pt, right=6pt, top=3pt, bottom=3pt,
  before skip=8pt, after skip=8pt}

\tikzset{
  card/.style={draw=fog, line width=0.6pt, rounded corners=2.2pt, fill=cardgray,
    align=center, font=\sffamily\scriptsize, inner sep=4pt},
  cardT/.style={card, fill=cardteal},
  cardA/.style={card, fill=cardamber},
  lane/.style={font=\sffamily\scriptsize\bfseries, text=slate, align=center},
  flow/.style={-{Stealth[length=2.4mm]}, line width=0.7pt, draw=slate,
    shorten <=1.3pt, shorten >=1.3pt, line cap=round},
  loop/.style={-{Stealth[length=2.2mm]}, line width=0.7pt, draw=amber,
    shorten <=1.3pt, shorten >=1.3pt, line cap=round},
  loopd/.style={-{Stealth[length=2.2mm]}, line width=0.7pt, draw=amber, dashed,
    shorten <=1.3pt, shorten >=1.3pt, line cap=round},
  note/.style={font=\sffamily\tiny, text=slate, align=center}
}

\begin{document}

\thispagestyle{plain}

\begin{center}
{\LARGE\bfseries Search over the Visual World:\\[3pt]
Persistent Visual Memory, Layered Indexes,\\[2pt] and Source-Grounded Evidence\par}
\vspace{11pt}
{\normalsize
Sankalp Nagaonkar \quad Rohit Garg \quad Ankit Raj \quad Ashish Choithani \quad Ashutosh Trivedi\par}
\vspace{5pt}
{\normalsize VideoDB\par}
{\small \texttt{\{sankalp, rohit, ankit, ashish, ashu\}@videodb.io}\par}
\vspace{6pt}
{\small Technical Report \,\textperiodcentered\, July 2026\footnote{Benchmark configurations and reproduction instructions are available at \url{https://github.com/video-db/search-over-the-visual-world}. The open-source \emph{Deep Search}
implementation of stateful retrieval discussed in \S\ref{sec:loops} is available
at \url{https://github.com/video-db/deepsearch}.}\par}
\end{center}

\vspace{8pt}

\begin{center}
\begin{minipage}{0.94\textwidth}
\begin{tcolorbox}[enhanced, colback=thesisbg, colframe=thesisframe,
  boxrule=0.5pt, arc=2.5pt, left=8pt, right=8pt, top=5pt, bottom=5pt]
\small\noindent\textbf{Thesis.}\; Search over the visual world cannot be
reduced to ranking video files. It requires infrastructure that turns archived
and continuously arriving media into persistent, provenance-bearing
understanding; organises that understanding through coexisting indexes;
selects bounded, task-specific context; and returns evidence that remains
playable at the original source. Search is the \emph{context interface}
between accumulated visual understanding and grounded action, and today it is a systems problem, not a training-time optimization.
\end{tcolorbox}
\end{minipage}
\end{center}

\vspace{2pt}

\begin{center}
\begin{minipage}{0.90\textwidth}
\small
\noindent\textbf{Abstract.}
Most video-retrieval systems assume a bounded corpus and return ranked files or timestamps. Agents operating over cameras, screens, streams, and archives face a different systems problem: observations arrive continuously; different models interpret them at different temporal granularities; useful context must be selected without replaying the complete visual record; and a result must remain connected to inspectable source evidence. We argue that search over such a
corpus is an infrastructure problem that cannot be reduced to ranking video
files. We develop a conceptual and formal model of \emph{search over the visual
world} built on analyzer-defined \emph{scenes}, persistent \emph{understanding
artifacts}, \emph{visual memory} as coexisting scene spaces over shared source
time, and capability-declared \emph{indexes}; we draw a strict distinction
between memory (everything retained), context (what is selected for a task), and
evidence (the source intervals that ground it). The \emph{VideoDB data format}
(VDB) realizes this conceptual model in production, and a typed search surface exposes it:
planned retrieval, a bounded stateful investigation mode, direct semantic,
structured, and aggregate access, and grounded synthesis. We contrast this model-agnostic infrastructure (segmentation, sampling, model choice, embeddings, and ranking all exposed as system decisions, with live streams as first-class sources) with video-native foundation models offered as fixed APIs. In a complete-system semantic-retrieval comparison against a commercial
video-native retrieval engine spanning 9{,}800+ natural-language queries over four public datasets, a
pipeline of general-purpose components, none trained end-to-end for video
retrieval, achieves higher macro-averaged Recall@1/@3/@10 (73.09/83.39/91.20
versus 65.75/77.13/89.10), while the baseline is higher at Recall@50 (96.42
versus 96.07). The results suggest that, today, retrieval quality over the
visual world is governed more by system design than by video-specific pretraining, and that visual-memory infrastructure can deliver it while
keeping source-grounded, playable evidence a first-class system object.
\end{minipage}
\end{center}

\vspace{6pt}

\section{Introduction}\label{sec:intro}

The visual world unfolds continuously. Cameras watch loading docks,
intersections, and operating rooms; screens record demonstrations, lectures, and
incidents; meetings and broadcasts stream for hours; archives accumulate years
of footage. AI systems are increasingly asked to act on this record: to notice,
recall, verify, and decide. Yet the software through which they reach it still
mostly exposes the visual world as isolated files and transient model outputs. A
multimodal model can analyze images and transcribe audio it is handed. An agent operating over months of
recordings and days of live streams must do something harder: preserve what has
already been observed, retrieve exactly the interval that matters, verify where
it came from, and consume the answer as media rather than as a paraphrase.
Recent long-video systems likewise route questions to selected temporal
segments, maintain complementary memory representations, or incrementally
construct retrievable state from streaming histories
\citep{kim2025salova,yeo2026worldmm,liang2026oasis,xie2026streamrag}.

Consider the questions such an agent must answer. \emph{When did the forklift
last enter bay~3, and was the dock door open?} \emph{Find every moment in
yesterday's launch stream where the demo failed.} \emph{Show the second time the
white car passes the intersection after 22:00.} Each question names an interval,
not a file. Temporal-grounding work formalizes this distinction by localizing short moments
within much longer sources \citep{soldan2022mad}. Each question also depends on several
kinds of understanding at once: objects, actions, speech, on-screen text. Each
ranges over a corpus that was still growing when the question was asked. And
each answer is useful only if it can be played and checked.

These questions expose six properties that jointly distinguish the visual world
from a document corpus. Each property imposes a systems obligation that no
model, however capable, discharges by itself
(Table~\ref{tab:obligations}).

\begin{table}[t]
\centering
\small
\renewcommand{\arraystretch}{1.12}
\begin{tabular}{@{}p{0.235\textwidth}p{0.335\textwidth}p{0.345\textwidth}@{}}
\toprule
\textbf{Property of the visual world} & \textbf{What it means} & \textbf{Systems obligation} \\
\midrule
\textbf{R1}\; Temporal & Meaning lives in intervals, order, and change, not in atomic items & Represent intervals and query-specific boundaries, not only files \\
\textbf{R2}\; Continuously changing & The corpus grows while being queried; a live stream has no final state & Understand and search live observations through the same abstraction as archives \\
\textbf{R3}\; Distributed & Sources span cameras, screens, streams, and recordings & Join heterogeneous observations on stable identity and shared source time \\
\textbf{R4}\; Larger than usable context & The corpus cannot be treated as one model context \citep{liu2024lost} & Select bounded, task-relevant context; never ship the corpus per request \\
\textbf{R5}\; Understood in pieces & No single model produces all the observations a task needs & Preserve several analyzer-defined views; force no canonical representation \\
\textbf{R6}\; Verifiable at the source & A conclusion an agent acts on must remain checkable against media & Retain provenance from every derived record down to a playable interval \\
\bottomrule
\end{tabular}
\caption{Properties of the visual world and the systems obligations they
impose.}
\label{tab:obligations}
\end{table}

None of this is removed by more capable models. Models do not, by themselves,
provide persistence, temporal alignment, indexing, retrieval, provenance,
permissions, live access, context selection, evidence delivery, or action.
These are obligations that arise before a model is invoked and after it
returns. Large-scale live-video systems likewise expose scheduling, resource,
quality, and processing-lag concerns as explicit systems obligations
\citep{zhang2017videostorm}. These obligations are the subject of this report.

\thesisbox{Search over the visual world cannot be reduced to ranking video
files. Its unit of retrieval is not the file but the \emph{scene}; its corpus is
not media but \emph{persistent visual memory}, the accumulated, provenance-bearing machine understanding of that media; and its output is not a
hit list but \emph{evidence}: source-grounded intervals that a person, a model,
or an agent can inspect, play, and act on. Supporting it requires infrastructure
that connects heterogeneous understanding, persistent memory, coexisting
indexes, bounded context selection, and directly consumable evidence, uniformly
over archived recordings and live streams.}

\begin{figure}[t]
\centering
\begin{tikzpicture}[
  chip/.style={draw=fog, line width=0.55pt, rounded corners=1.8pt, fill=white,
    align=center, font=\sffamily\tiny, inner sep=2.6pt},
  band/.style={font=\sffamily\tiny\bfseries, text=slate!75}]

  \node[band] at (1.00,3.34) {SOURCE};
  \node[band] at (4.10,3.34) {UNDERSTAND};
  \node[band] at (7.85,3.34) {PERSIST};
  \node[band] at (11.55,3.34) {RETRIEVE};
  \node[band] at (14.65,3.34) {GROUND \& ACT};

  \foreach \y/\t in {2.55/Cameras, 1.85/Screens, 1.15/Live Streams, 0.45/Archives}
    { \node[chip, minimum width=1.55cm, minimum height=0.46cm] at (1.00,\y) {\t}; }

  \draw[draw=fog, line width=0.6pt, rounded corners=2.5pt, fill=cardgray]
    (2.72,-0.10) rectangle (5.48,3.10);
  \node[font=\sffamily\scriptsize\bfseries, text=ink] at (4.10,2.86) {Understanding};
  \node[font=\sffamily\tiny, text=slate] at (4.10,2.58) {analyzer portfolio};
  \foreach \y/\t in {2.14/{\textbf{ASR}\; transcript}, 1.68/{\textbf{VLM}\; describe},
                     1.22/{\textbf{Detector}\; objects}, 0.76/{\textbf{OCR}\; screen text},
                     0.30/{\textbf{Domain}\; events}}
    { \node[chip, minimum width=2.30cm, minimum height=0.40cm] at (4.10,\y) {\t}; }

  \draw[draw=teal!55, line width=0.6pt, rounded corners=2.5pt, fill=cardteal]
    (6.30,-0.10) rectangle (9.40,3.10);
  \node[font=\sffamily\scriptsize\bfseries, text=ink] at (7.85,2.86) {Visual Memory};
  \node[font=\sffamily\tiny, text=slate] at (7.85,2.58) {scenes, artifacts, provenance};
  \foreach \s/\e in {6.55/7.20, 7.35/8.20, 8.35/9.15}
    { \draw[fill=rowa!16, draw=rowa, line width=0.5pt] (\s,2.12) rectangle (\e,2.32); }
  \foreach \s/\e in {6.55/7.60, 7.60/8.45, 8.45/9.15}
    { \draw[fill=rowb!14, draw=rowb, line width=0.5pt] (\s,1.84) rectangle (\e,2.04); }
  \foreach \i in {0,...,5}
    { \draw[fill=rowc!14, draw=rowc, line width=0.45pt]
      ($(6.55+\i*0.44,1.56)$) rectangle ($(6.89+\i*0.44,1.76)$); }
  \draw[line width=0.55pt, draw=slate, -{Stealth[length=1.8mm]}] (6.55,1.32) -- (9.15,1.32);
  \node[font=\sffamily\tiny, text=slate] at (7.85,1.10) {Aligned on source time};
  \foreach \x/\t in {6.87/Semantic, 7.85/Structured, 8.83/Aggregate}
    { \node[chip, minimum width=0.90cm, minimum height=0.36cm] at (\x,0.66) {\t}; }
  \node[font=\sffamily\tiny, text=slate] at (7.85,0.26) {Capability-declared index layers};

  \draw[draw=fog, line width=0.6pt, rounded corners=2.5pt, fill=cardgray]
    (10.15,-0.10) rectangle (12.95,3.10);
  \node[font=\sffamily\scriptsize\bfseries, text=ink] at (11.55,2.86) {Search Surface};
  \node[chip, minimum width=2.50cm, minimum height=0.52cm] at (11.55,2.30)
    {\textbf{\textit{Search}}\\ planned retrieval};
  \node[chip, minimum width=2.50cm, minimum height=0.52cm, fill=cardamber] at (11.55,1.64)
    {\textbf{\textit{Deep Search}}\\ stateful refinement};
  \node[chip, minimum width=2.50cm, minimum height=0.40cm] at (11.55,1.06)
    {\textit{Semantic, Query, Aggregate}};
  \node[chip, minimum width=2.50cm, minimum height=0.52cm] at (11.55,0.44)
    {\textbf{\textit{Ask}}\\ grounded synthesis};

  \node[card, text width=2.05cm, minimum height=0.66cm] (ctx) at (14.65,2.44)
    {\textbf{Context}\\[0.5pt] {\tiny bounded, task-relevant}};
  \node[cardT, text width=2.05cm, minimum height=0.66cm] (evi) at (14.65,1.30)
    {\textbf{Evidence}\\[0.5pt] {\tiny playable source intervals}};
  \node[cardA, text width=2.05cm, minimum height=0.44cm] (agent) at (14.65,0.22)
    {\textbf{Agent Acts}};

  \foreach \y in {2.55,1.85,1.15,0.45} { \draw[flow] (1.80,\y) -- (2.72,\y); }
  \draw[flow, line width=1.0pt] (5.48,1.50) -- (6.30,1.50);
  \draw[flow, line width=1.0pt] (9.40,1.50) -- (10.15,1.50);
  \draw[flow] (12.95,2.44) -- (ctx.west);
  \draw[flow] (ctx.south) -- (evi.north);
  \draw[flow] (evi.south) -- (agent.north);

  \draw[loop] (11.95,-0.10) to[out=-70,in=-110,looseness=2.2]
    node[below=1pt, note]{Query-time refinement (\S\ref{sec:loops})} (11.15,-0.10);
  \draw[loopd] (agent.south) to[out=-125,in=-55,looseness=0.50]
    node[below=0.5pt, note, pos=0.55]{Representation refinement: Re-understand and Re-index (\S\ref{sec:loops})}
    (4.60,-0.10);
\end{tikzpicture}
\caption{Search over the visual world. Sources feed a portfolio of analyzers; each row is a choice of model, segmentation, sampling, prompt, and schema, not a commitment fixed at training time. The VDB data format
(\S\ref{sec:vdm}) persists their outputs as visual memory: scenes and artifacts aligned on
source time, carrying provenance, with capability-declared index layers over
them. A typed search surface converts memory into bounded context and playable
evidence for an agent. Two loops close the system: query-time refinement
revises the plan within an investigation, and representation refinement
directs re-understanding and re-indexing of memory itself.}
\label{fig:pipeline}
\end{figure}
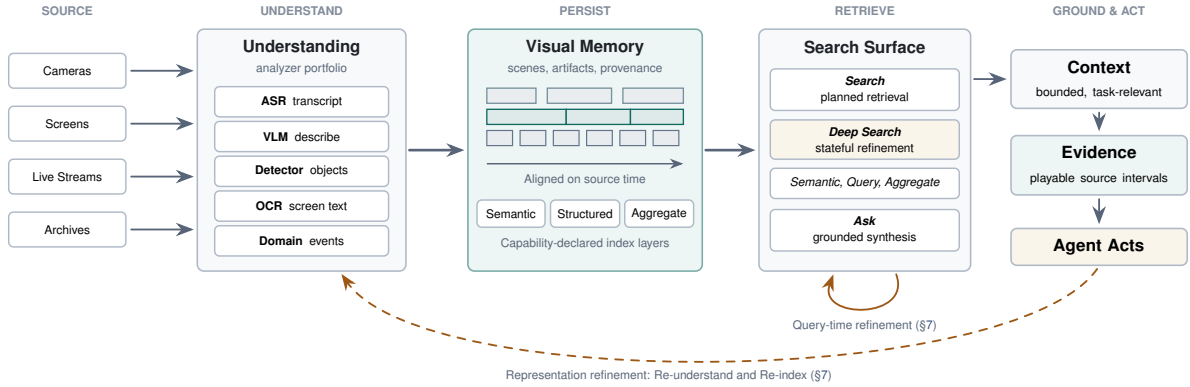

The argument proceeds as a chain of definitions with one feedback path
(Figure~\ref{fig:pipeline}). \emph{Models produce understanding. A visual data
format turns understanding into persistent memory. Indexes make memory
searchable. Search selects the memory relevant to a task. The selected memory
becomes context. Source-linked context becomes evidence.} In this account,
search is the \emph{context interface} of the visual world: the mechanism that
translates accumulated visual understanding into a compact, task-conditioned
representation while preserving the route back to the original media. Search
plays a second role as well: it is the probe that reveals whether the current
representations of memory are adequate, and thereby drives their revision
(\S\ref{sec:loops}).

We instantiate and evaluate these ideas in VideoDB, a production visual data
infrastructure \citep{videodb2026docs}. In the vocabulary of this report: search over the visual world
is the scientific problem; visual data infrastructure is the systems category;
the \emph{VideoDB data format} (VDB) is the mechanism; persistent visual memory
is the resulting capability; search is the interface through which an agent
converts memory into task-specific context; streamable evidence is the
source-grounded output; and VideoDB is the reference implementation and the
evaluated system. The system is the vehicle of the paper, not its object.

\paragraph{Contributions}
This report makes four contributions.
\begin{enumerate}[leftmargin=1.6em, itemsep=1.5pt, topsep=3pt]
  \item A problem formulation and conceptual model of search over the visual
  world: analyzer-defined scenes, persistent understanding artifacts, visual
  memory as coexisting scene spaces over shared source time, and a strict
  memory--context--evidence distinction (\S\ref{sec:model}).
  \item The VideoDB data format (VDB): a logical data format binding source
  identity, source time, artifacts, provenance, and capability-declared
  indexes, with understanding separated from access so that representations
  can be re-derived without re-analysis, and tightly coupled to a streaming
  engine that realizes any logical selection, at millisecond granularity,
  as directly playable media (\S\ref{sec:vdm}--\S\ref{sec:evidence}).
  \item A typed search surface over visual memory (planned retrieval, a bounded stateful investigation mode, direct semantic/structured/aggregate contracts, and grounded synthesis), together with the distinction between
  \emph{indexed} and \emph{resolved} scenes and a two-loop account of
  closed-loop visual search (\S\ref{sec:search}--\S\ref{sec:loops}).
  \item An architectural comparison of video-native model APIs with
  model-agnostic visual data infrastructure, and a controlled complete-system
  semantic-retrieval evaluation against a commercial video-native retrieval
  engine using 9{,}800+
  natural-language queries from four public datasets, with explicitly scoped
  claims (\S\ref{sec:architectures}--\S\ref{sec:eval}).
\end{enumerate}

\section{The Shape of the Problem}\label{sec:problem}

\subsection{From corpus to world}

Classical retrieval rests on assumptions that quietly fail here. A document
corpus has a stable atomic unit (the document), one dominant representation
(its text), an update model of insertion, and results a person can read
directly. Video--text retrieval inherits the same shape: encode clip and query
into one embedding space and rank
\citep{radford2021learning,bain2021frozen,ni2022xclip}. This is effective for
what it models, and we use such representations throughout. But as an account
of search over the visual world it fails three ways at once. The \emph{unit} is
wrong: answers are intervals within and across sources, not files
(R1). The \emph{representation} is wrong: no single embedding space encodes
counts, on-screen text, identities, procedures, and domain events at once, and
different tasks need these at different fidelities and costs (R5). The
\emph{lifecycle} is wrong: the corpus grows while being queried (R2), model
outputs evaporate unless persisted, and a ranked identifier is not something an
agent can play or verify (R6).

Retrieval-augmented generation established the complementary lesson for text:
models benefit from selective external context rather than from ingesting the
corpus \citep{lewis2020retrieval}, and long-context studies explain why bounded,
well-chosen context beats indiscriminate stuffing \citep{liu2024lost}. Recent
work extends retrieval augmentation to video corpora \citep{jeong2025videorag},
and agent-memory systems demonstrate persistent stores mediated by retrieval
\citep{park2023generative,packer2023memgpt}; but their memory objects are text records. What remains open is the question those lines converge on: \emph{what
is the retrieved unit of visual experience, and what infrastructure must exist
for it to be produced, kept, found, bounded, and believed?}

\subsection{The missing layer between models and media}

Media containers and streaming protocols preserve encoded content and timing.
Semantic observations, when they are kept at all, live in application-specific
side stores. The result is a familiar production pathology: a second
application repeats decoding and inference, writes a different schema, and
implements its own path back to the source; observations produced for one
pipeline cannot be reused by another; and nothing ties a derived row to the
seconds of source video that justify it. The problem is not that a media file
lacks a timeline. The problem is that stable source identity, source time,
derived observations, retrieval capabilities, and evidence delivery never meet
in one reusable machine interface.

Three database lessons shape the layer that is missing. First, derived results
should be reusable across materializations rather than owned by one query path: the materialized-view principle \citep{gupta1995maintenance}. Second,
every result should retain lineage to its source \citep{buneman2001why};
without lineage, verification (R6) degenerates into trust. Third, no single
engine or representation serves all workloads \citep{stonebraker2005one}; the
consequence for visual data is not many databases but many coexisting access
paths over one shared substrate. Ad hoc glue between models and media is also
exactly the entanglement that accumulates as technical debt in production ML
systems \citep{sculley2015hidden}.

\subsection{What existing systems solve}

Prior video systems address important stages of this lifecycle. VideoStorm
manages resource, quality, and lag tradeoffs for live analytics
\citep{zhang2017videostorm}, while Scanner schedules raster computation over
large video collections \citep{poms2018scanner}. Boggart builds query-independent
representations for later heterogeneous analytics \citep{agarwal2023boggart},
and EVA materializes expensive video-analysis outputs for reuse across
subsequent queries \citep{xu2022eva}. Rekall composes spatiotemporal labels into
event queries \citep{fu2019rekall}. TASTI builds semantic indexes that
accelerate ML-based queries over unstructured data \citep{kang2022tasti}. VIVA
supports interactive, relational video analytics with model selection
\citep{kang2022viva}. V2V demonstrates that query results can themselves be
synthesized as video \citep{winecki2024v2v}. Together these systems cover
computation, event algebra, reuse, index acceleration, analytics, and
presentation. They primarily specialize in particular lifecycle stages rather
than exposing as their central contract the connective layer studied here:
heterogeneous, open-schema observations persisted with provenance; multiple
capability-declared indexes coexisting over the same sources; and evidence that
remains playable media. That connective layer is what this report formalizes
and what VideoDB implements.

\section{A Conceptual Model of Search over the Visual World}\label{sec:model}

We now define the objects of study. Table~\ref{tab:vocab} summarizes the
vocabulary; Figure~\ref{fig:scenes} illustrates it over a small collection.

\begin{table}[t]
\centering
\small
\begin{tabular}{@{}p{0.185\textwidth}p{0.765\textwidth}@{}}
\toprule
\textbf{Term} & \textbf{Definition} \\
\midrule
\term{understanding} & Production of temporally grounded observations from visual media by a compatible analytical process. \\
\term{analyzer} & A concrete understanding process: a model with a segmentation policy and configuration (sampling, prompt, schema). \\
\term{scene} & An analyzer-defined, source-aligned temporal unit selected for understanding and indexing. \\
\term{artifact} & A persistent, model-derived observation attached to a scene, carrying provenance. \\
\term{visual memory} & The persistent collection of source-aligned scenes, artifacts, indexes, and provenance available for later retrieval. \\
\term{index} & A materialized access path over selected artifacts, exposing declared capabilities. \\
\term{search} & Selection and ranking of relevant scenes from visual memory according to an intent and optional constraints. \\
\term{context} & The bounded, task-relevant subset of visual memory selected for a model or agent. \\
\term{evidence} & The original, source-grounded interval supporting a result, conclusion, or action, consumable as media. \\
\term{stream} & The playable realization of evidence: any interval, or ordered composition of intervals across sources, generated on demand by the streaming engine directly from source-time coordinates. \\
\term{indexed scene} & An analyzer-defined temporal unit created to support understanding and retrieval. \\
\term{resolved scene} & A query-specific evidence interval produced after reasoning over retrieved candidates. \\
\bottomrule
\end{tabular}
\caption{The vocabulary of search over the visual world. Memory is everything
retained; context is what is selected for the present task; evidence is the
source material that grounds it; a stream is evidence made playable, on
demand.}
\label{tab:vocab}
\end{table}

\subsection{Sources and time}

A \emph{source} is an archived video or a live stream. Every source has two
invariants: a stable identity, and a \emph{source-time axis}: the clock of the media itself, starting at zero. For an archived recording the axis has a fixed
extent; for a live stream it is still growing, and there is no final state
(R2). Source time is the coordinate system in which every observation about a
source is expressed. It is what lets observations produced by different
processes, at different times, by teams that have never coordinated, meet again
later on the same timeline.

\subsection{Understanding, analyzers, and scenes}

\begin{definition}[Understanding]\label{def:understanding}
Understanding is the production of temporally grounded observations from
visual media by a compatible analytical process.
\end{definition}

Understanding is deliberately plural (R5). It may be produced by classical
computer-vision models; object detectors and trackers; OCR and speech systems;
specialised domain models; small or medium open-weight vision--language models;
hosted multimodal models; or frontier models, including video-native foundation models, which enter this framework as one member among many
(\S\ref{sec:architectures}). These processes differ by orders of magnitude in
cost and latency and by task in reliability. Visual understanding is not the
output of one large model; it is a portfolio, and the portfolio changes over
time.

\begin{definition}[Analyzer]\label{def:analyzer}
An analyzer is a concrete understanding process: a model, a temporal
segmentation policy, and a configuration: its frame-selection and sampling strategy, prompt, parameters, and output schema.
\end{definition}

Every element of an analyzer is a choice. Which model reads the frames, how
the timeline is cut, how densely frames are sampled, what the model is asked,
and what shape its answer takes are decisions made per analyzer, per workload, not properties fixed once for the whole system. Holding these
choices open, rather than freezing them inside a trained artifact, is the
design decision from which the rest of this report follows.

\begin{definition}[Scene]\label{def:scene}
A scene is an analyzer-defined, source-aligned temporal unit selected for
understanding and indexing.
\end{definition}

\begin{figure}[t]
\centering
\begin{tikzpicture}[x=0.86cm]
  \draw[line width=0.7pt, draw=slate, -{Stealth[length=2.6mm]}] (0,0) -- (16.0,0);
  \node[note, anchor=west] at (14.55,-0.34) {source time $T_v$};
  \foreach \t [evaluate=\t as \lab using int(\t*10)] in {0,2,...,14}
    { \draw[draw=fog] (\t,0) -- (\t,0.09); \node[note] at (\t,-0.29) {\scriptsize \lab s}; }

  \node[lane, anchor=east] at (-0.28,0.68) {Source A\\Speech};
  \foreach \s/\e in {0.3/1.9, 2.3/4.6, 5.4/6.1, 7.0/9.8, 10.9/13.2, 13.8/15.4}
    { \draw[fill=rowa!14, draw=rowa, line width=0.6pt, rounded corners=1pt] (\s,0.42) rectangle (\e,0.94); }

  \node[lane, anchor=east] at (-0.28,1.58) {Source A\\Shot/VLM};
  \foreach \s/\e in {0.0/2.8, 2.8/5.2, 5.2/7.4, 7.4/8.6, 8.6/11.5, 11.5/15.6}
    { \draw[fill=rowb!12, draw=rowb, line width=0.6pt, rounded corners=1pt] (\s,1.32) rectangle (\e,1.84); }

  \node[lane, anchor=east] at (-0.28,2.48) {Source A\\Detector};
  \foreach \i in {0,...,18} { \draw[fill=rowc!12, draw=rowc, line width=0.5pt]
    ($(\i*0.84,2.22)$) rectangle ($(\i*0.84+0.68,2.74)$); }

  \node[lane, anchor=east] at (-0.28,3.38) {Source B\\Domain};
  \foreach \s/\e in {1.2/3.6, 5.1/7.0, 10.0/12.9}
    { \draw[fill=rowd!14, draw=rowd, line width=0.6pt, rounded corners=1pt] (\s,3.12) rectangle (\e,3.64); }

  \draw[draw=amber, line width=1.1pt, rounded corners=1pt] (7.0,0.42) rectangle (9.8,0.94);
  \draw[draw=amber, line width=1.1pt, rounded corners=1pt] (8.6,1.32) rectangle (11.5,1.84);
  \draw[draw=amber, line width=1.1pt, rounded corners=1pt] (10.0,3.12) rectangle (12.9,3.64);

  \draw[dashed, draw=teal, line width=0.8pt] (8.32,2.00) -- (8.32,-0.52);
  \draw[dashed, draw=rowd, line width=0.8pt] (11.62,3.90) -- (11.62,-0.52);

  \node[lane, anchor=east] at (-0.28,-1.55) {VDB timeline\\program};
  \draw[fill=rowb!16, draw=teal, line width=0.8pt, rounded corners=1.5pt]
    (2.0,-1.86) rectangle (7.9,-1.24);
  \node[font=\sffamily\scriptsize, text=ink] at (4.95,-1.55)
    {Source A: arbitrary $[t_0,\,t_1)$};
  \draw[fill=rowd!16, draw=rowd, line width=0.8pt, rounded corners=1.5pt]
    (7.9,-1.86) rectangle (12.4,-1.24);
  \node[font=\sffamily\scriptsize, text=ink] at (10.15,-1.55)
    {Source B: arbitrary $[t_2,\,t_3)$};
  \node[note, anchor=west] at (12.65,-1.55) {new HLS stream};

  \draw[-{Stealth[length=2.2mm]}, line width=0.8pt, draw=teal]
    (8.32,-0.52) to[out=-90,in=55] (5.7,-1.20);
  \draw[-{Stealth[length=2.2mm]}, line width=0.8pt, draw=rowd]
    (11.62,-0.52) to[out=-90,in=70] (10.4,-1.20);
\end{tikzpicture}
\caption{Coexisting scene spaces over a two-source collection, and the stream
they resolve into. Analyzers segment each source differently: speech by
utterance, a vision--language model by shot, a detector by fixed stride, a
domain analyzer by events; every scene is anchored to shared source time.
For a query, candidates (outlined) may come from several spaces, in several
sources, with non-identical boundaries, and the resolved boundaries (dashed)
are constructed to fit the question, coinciding with no stored boundary
(\S\ref{sec:resolved}). \emph{Bottom:} a VDB timeline program concatenates
arbitrary spans $[t_0,t_1)$ of source A and $[t_2,t_3)$ of source B, and the
streaming engine delivers the program as one new HLS stream, generated on
demand (\S\ref{sec:evidence}). With fixed segments and fixed files off the
read path, the whole visual world collapses into flexible start--end
coordinates over a collection: a data system, not a file system.}
\label{fig:scenes}
\end{figure}
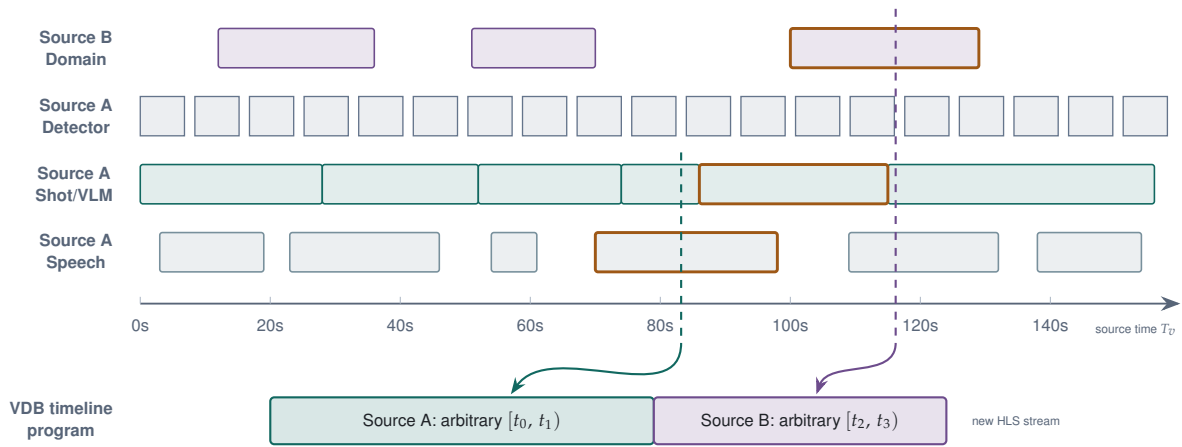

We do not impose a universal scene segmentation. Online event-boundary research
likewise treats segmentation for a live source as a causal operation over
present and past frames rather than one that assumes access to the finished
video \citep{jung2025onlinegebd}. A speech system segments by utterance; a shot
detector by visual cut; a sampling detector by fixed stride; a domain analyzer
by domain events; a summarizer by narrative arc. Applying an
analyzer to a source yields that analyzer's \emph{scene space}: the set of
scenes it produced, each recording four things: which source, which interval
of source time, what was observed there, and the provenance of the
observation. Distinct analyzers produce different, overlapping scene spaces
over the same source: different boundaries, sampling frequencies, temporal
resolutions, models, prompts, and schemas. Nothing requires the spaces to
agree, and useful ones typically do not (Figure~\ref{fig:scenes}). What
unifies them is not shared boundaries but the shared source-time axis.

\begin{definition}[Understanding artifact]\label{def:artifact}
An understanding artifact is a persistent, model-derived observation associated
with a scene and its provenance.
\end{definition}

Artifacts include descriptions, embeddings, transcript spans, detected objects
and counts, tracks, OCR text, actions, locations, emotions, measurements, and
domain-specific structures. The payload schema is open: these are examples, not
an ontology. An artifact's provenance record states which analyzer, model,
prompt, schema, and version produced the observation, and when. This follows
database accounts in which lineage is retained and propagated through derived
results \citep{buneman2001why,green2007provenance}. Provenance is what makes an
artifact citable, comparable with artifacts from other analyzers, and safely
replaceable. It is also what makes understanding
\emph{inspectable}: artifacts are data a user can read, export, and build on,
not internal state hidden behind a retrieval endpoint.

\subsection{Visual memory}

\begin{definition}[Visual memory]\label{def:memory}
Visual memory is the persistent collection of source-aligned scenes,
understanding artifacts, indexes, and provenance available for later
retrieval.
\end{definition}

Visual memory is the union of every scene space produced over every source in
a collection, together with the indexes built over them and the provenance
that explains them. The term is operational, not an analogy to human memory:
it denotes durable, source-aligned observations plus the access structures
that let later requests recover bounded evidence. Model-level long-video work
also uses complementary memory representations for query-dependent reasoning
\citep{yeo2026worldmm}; those inference-time memories are related motivation,
not equivalents of the independently managed visual memory defined here. Two
properties matter.
Visual memory is the \emph{complete retained superset}; it is never what is handed to a model. And it is \emph{append-mostly and revisable}: new analyzers
add coexisting views, and provenance allows old views to be replaced or
retired deliberately rather than silently.

\subsection{Indexes}

\begin{definition}[Index]\label{def:index}
An index is a materialized access path over selected understanding artifacts,
exposing declared capabilities.
\end{definition}

An index is derived from a chosen set of stored artifacts and declares what it
can do: semantic retrieval over embedded fields, structured filters over typed
fields, aggregation and metrics over grouped fields, computed properties,
ranking, and domain-specific access. Prior video systems show that expensive
analysis outputs can be materialized for reuse across later queries
\citep{xu2022eva}. Each index is a searchable
\emph{understanding layer} over the sources it covers, and several layers
routinely coexist over the same source even when their scene boundaries
differ. Because artifacts are exposed rather than hidden, indexing is
something users do, not only something done to them: an application can read
the stored understanding and index it its own way: its own embedding model, its own computed fields, its own layer composition. Two consequences follow
from keeping stored understanding strictly separate from the indexes over it.
A new index can be built from existing artifacts \emph{without re-running the
source-understanding model}: derivation reads artifacts; it never re-invokes
the analyzer. And capability declaration lets a planner reason over the
access paths that actually exist instead of inventing unsupported operations
(\S\ref{sec:search}). Related database systems optimize among alternative
semantic operators under explicit quality, cost, and latency constraints
\citep{russo2026abacus}, and combine vector traversal with relational filters
and joins \citep{zhang2023vbase}; these works motivate explicit capabilities
without defining the visual data format used here. Computed properties make the point concrete: a detector
artifact materialised as a numeric \texttt{person\_count} lets \qtext{scenes
where more than two people are present} evaluate as a typed filter over stored
understanding, rather than as a fresh model pass over the corpus.

One further property distinguishes these indexes from conventional metadata
search. Because every indexed scene carries source-time coordinates in the
same format the streaming engine reads (\S\ref{sec:vdm}), an index hit is
not a pointer that some downstream export job must redeem: it terminates in
an interval that is directly playable, and any set of hits, within or across
sources, is directly composable into a single stream
(\S\ref{sec:evidence}). An index built over stored understanding is
therefore also, automatically, an index into deliverable media.

\subsection{Search, context, and evidence}

\begin{definition}[Search]\label{def:search}
Search is the selection and ranking of relevant scenes from visual memory
according to an intent and optional constraints.
\end{definition}

Search takes an intent and optional constraints and proceeds in two stages.
First, \emph{candidate retrieval}: one or more index layers are consulted, and
each contributes candidate scenes. Second, \emph{reasoning over candidates}:
the candidates are validated against the intent, ranked, and resolved into
results. Candidates drawn from different layers need not share scene
boundaries; they are combined on the source-time axis.

\begin{definition}[Context]\label{def:context}
Context is the bounded, task-relevant subset of visual memory selected for a
model or agent.
\end{definition}

Context is assembled, not found. Assembling it involves four decisions:
\emph{selection} (which intervals matter for this task), \emph{hydration}
(which stored fields ride along with them), \emph{representation} (the cheapest faithful form: a transcript span, a description, a handful of frames, or the playable interval itself), and \emph{budget} (how much the
consuming model can use well \citealp{liu2024lost}). The same visual memory
yields different contexts for different tasks, and none of them changes the
memory: context is disposable and rebuilt per task, while memory is durable.
An agent does not carry the visual world with it; it carries the ability to
retrieve exactly the part of the world the present task needs. We call the
resulting property \emph{context efficiency}: the system returns the representation an information need requires (scenes, rows, fields, or grounded text) rather than passing the visual record to a model. This report
treats context efficiency as a design property of the interface; measuring it
in model calls, visual tokens, and cost is future benchmark work
(\S\ref{sec:limits}).

\begin{definition}[Evidence]\label{def:evidence}
Evidence is the original, source-grounded visual interval supporting a search
result, conclusion, or action, retained in directly consumable form.
\end{definition}

Evidence is not a generated answer, a similarity score, a file identifier, a
textual citation, or a timestamp without an operational path to the media. An
evidence item names its source and its interval on the source-time axis,
carries the fields attached to it, and can be produced as media on demand
(\S\ref{sec:evidence}).

\emph{Produced as media} is meant literally, and it adds a further term to
the vocabulary. A \term{stream} is the playable realization of evidence: any
interval, or ordered composition of intervals across sources, delivered on
demand by the streaming engine coupled to the data format of
\S\ref{sec:vdm}. The timestamps this vocabulary traffics in are therefore not
bookkeeping about media kept elsewhere; they are playable coordinates.
Nothing in the delivery path requires a stored segment boundary or an
intermediate file: the engine can begin at any point of any source, end at
any other, and compose what lies between, together with overlays from other
modalities, into media a person or an agent watches seconds after asking
(\S\ref{sec:evidence}). The distinction between \emph{finding} a moment and
\emph{having} it as watchable media disappears. These distinctions anchor
everything that follows:

\thesisbox{Memory is everything retained. Context is what is selected for the
present task. Evidence is the source material that grounds it. A stream is
evidence made playable, on demand.}

\section{The VideoDB Data Format (VDB)}\label{sec:vdm}

The conceptual model becomes infrastructure through a data format. The
\emph{VideoDB data format} (henceforth \emph{VDB}, or the \emph{VDB format})
is the durable representation in which the visual world is stored, indexed,
and, just as essentially, \emph{delivered}: once a source enters VDB, every
later operation, from retrieval to composition to playback, is expressed
over the format rather than over media files. For each source, VDB maintains
one persistent cell that binds five things: the source's stable identity;
its source-time axis; the scene spaces and artifacts accumulated by every
analyzer that has run over it; the indexes derived from those artifacts; and
the provenance connecting each derived record to the analyzer, model,
schema, and moment that produced it. Collections group cells and may carry
collection-scoped indexes over many sources. Search contracts operate over
cells and collections and return evidence views; results are never folded
back into the cell as opaque state. The separation of logical video
operations from physical representation has precedent in video storage
systems \citep{haynes2021vss}; VDB extends that separation to understanding
artifacts, indexes, provenance, and evidence.

\paragraph{A logical data format}
VDB is a \emph{logical} data format: it standardizes what the records of the
visual world are and how they connect, while prescribing no codec, no
container, and no byte layout. The distinction is the classical one between
a logical model and its physical realization \citep{codd1970relational}, and
the nearest contemporary precedent is the open table format of analytic data
systems \citep{armbrust2020delta}: a logical specification layered over
ordinary physical files, made operational by engines, through which files
begin to behave like a database. VDB stands in the same relation to media.
Sources remain ordinarily encoded media, and the format's contribution is
the set of connections it makes durable: identity to time, time to scenes,
scenes to artifacts, artifacts to provenance, artifacts to indexes, and
indexes to evidence. What makes the format
actionable rather than descriptive is its second, coupled half: a
proprietary streaming \emph{editing} engine that reads VDB directly. We
characterize the engine by its observable contract only; the mechanism that
realizes the contract is deliberately out of scope for this report. The
contract is \emph{temporal data independence}: an application addresses
media purely in logical coordinates (this source or set of sources, this
span of source time, from any timestamp to any timestamp, at millisecond
granularity) and receives a conforming, immediately playable stream. No
stored file boundary, segment boundary, or packaging decision is observable
at the interface, and none constrains what can be addressed. Delivery
arrives as a standard HLS (M3U8) manifest \citep{pantos2017http}, playable
on commodity players, including Apple devices, with no special client
support; overlays from other modalities (image, audio, text, captions) are
declared over the same source-time axis and arrive composed into the
delivered stream \citep{videodb2026timeline}. Generation is an
interactive-latency operation, not a batch render
(Appendix~\ref{app:operational}, \S\ref{sec:evidence}).

\paragraph{An alternative to file-based pipelines}
This coupling is what retires the media file as the working medium
\citep{videodb2026mp4}. In file-based pipelines, every compilation, merge,
edit, or overlay is a render job that decodes containers and encodes a new
one; each derived MP4 is opaque to the next tool, and the archive fills with
artifacts whose relationship to their sources is maintained by convention.
Under VDB the same operations are declarations over logical coordinates,
exposed to developers as a declarative \emph{timeline} of clips and tracks
over source time \citep{videodb2026timeline} and evaluated at request time
by the streaming engine; no intermediate file is created or touched,
provenance survives because composition never leaves the format, and an MP4
or HLS rendition remains available as an \emph{export} at the edge for
interoperability, not as the substrate of work. Media becomes
something the system computes over and responds with, the way a database
responds to queries, rather than something it renders. Timestamps in VDB are
playable coordinates, and \S\ref{sec:evidence} develops the consequence:
search whose results are watchable the moment they are named.

\paragraph{P1: Understanding is separated from access}
Artifacts persist independently of the indexes over them
(\S\ref{sec:model}). This is the materialized-view principle applied to
machine understanding \citep{gupta1995maintenance,xu2022eva}: analysis is the
slowly changing investment; access paths are derived and plural. A new
index over existing artifacts requires no re-analysis; a replacement analyzer
is useful exactly when it emits an artifact schema compatible with the intended
access paths.

\paragraph{P2: Capabilities are declared}
Each index declares what it supports: semantic fields expose vector retrieval;
queryable fields expose typed conditions; aggregatable fields expose grouping
and metrics; stored fields can be hydrated into results. Declaration keeps the
artifact model open-ended (any collection can advertise its own fields and
types) while keeping planning honest: a planner selects among real access
paths rather than assuming a universal one \citep{stonebraker2005one}.

\paragraph{P3: Schemas are open}
Payloads and schemas are application-defined. Descriptions, transcripts,
detections, OCR, measurements, and domain structures are examples, not a closed
ontology; future analyzers add views without migrating old ones.

\paragraph{P4: Source time is the join key}
Every scene in every space is anchored to $T_v$. Temporal combination uses
source and aligned interval keys, so observations from analyzers that have
never heard of each other can be intersected, sequenced, and compared at
retrieval time (R1, R3).

\paragraph{P5: Provenance is first-class}
Every derived record can answer: which analyzer, which model, which prompt and
schema, over which interval, when \citep{buneman2001why}. Provenance is what
allows layered memory to be audited, selectively rebuilt, and safely retired,
and it survives into results so that evidence remains attributable.

\paragraph{P6: One format for archives and streams}
Nothing in the cell assumes a finished source. For a live stream, the
source-time axis grows; analyzers with time-based segmentation append scenes
as media arrives; indexes ingest them incrementally; identity and provenance
are unchanged (R2). Archived and live media differ in the state of one axis,
not in kind; this is why search, alerting, and evidence work over a stream
that is still happening, and why an agent can be handed playable evidence of
an event from moments ago (\S\ref{sec:evidence}) \citep{videodb2026docs}.

\paragraph{P7: Every selection is streamable}
Any temporal selection expressible in the format (a stored scene, a
query-resolved interval, an ordered set of intervals spanning many sources
of a collection) has a playable realization generated on demand by the
streaming engine, together with any overlays declared over it. No pre-cut
segment inventory and no per-request export job stand between a selection
and its playback: the delivery path reads the format, not files
(\S\ref{sec:evidence}).

The concrete analyzer, segmentation, schema, indexing, and retrieval
configuration used in the evaluation is reported in \S\ref{sec:benchmark-config};
the complete dataset-specific prompts appear in
Appendix~\ref{app:benchmark-prompts}.

\section{Evidence as a Stream}\label{sec:evidence}

The system object that closes the path from retrieval to action is the
evidence stream, and its delivery properties are benchmarked directly. Every
temporal result names its source and its interval; the streaming engine
produces playback of exactly that interval, and an ordered set of resolved
scenes, possibly spanning several sources of a collection, is composed into
a single view delivered through standard manifest-based streaming
\citep{videodb2026docs,pantos2017http}, continuing prior work that treats
video itself as a query result \citep{winecki2024v2v}. Because the stream is
constructed at request time from logical coordinates (P7), it is
\emph{boundary-free}: it can begin at any millisecond of any source,
archived or still live, and end at any other, with no stored segmentation
observable at the interface (Figure~\ref{fig:scenes}). It is also
\emph{immediate}. Appendix~\ref{app:operational} reports observed median
generation latency for playable evidence streams assembled from one source and
from five sources, at requested durations from five seconds to ten minutes.
Median generation remains below one second in every measured condition:
433--728~ms for single-source streams and 433--994~ms for five-source streams.
We are not aware of another production video-search platform that documents an
equivalent primitive for real-time cross-source, overlay-bearing composition. Verification of a retrieval result
thereby changes from an act of trust into an act of playback.

\section{Search over Visual Memory}\label{sec:search}

\begin{figure}[t]
\centering
\begin{tikzpicture}[
  chip/.style={draw=fog, line width=0.55pt, rounded corners=1.8pt, fill=white,
    align=center, font=\sffamily\scriptsize, inner sep=3.2pt},
  modetag/.style={rounded corners=1.8pt, align=center,
    font=\sffamily\scriptsize\bfseries, text=white, inner sep=3.4pt},
  outchip/.style={chip, fill=cardteal, draw=teal!45},
  panel/.style={draw=fog, line width=0.6pt, rounded corners=3pt, fill=cardgray},
  ptitle/.style={font=\sffamily\tiny\itshape, text=slate},
  tinyflow/.style={-{Stealth[length=2.0mm]}, line width=0.6pt, draw=slate,
    shorten <=1.2pt, shorten >=1.2pt, line cap=round}]

  \node[lane, anchor=center] at (0.68,4.00) {HIGH-LEVEL\\SEARCH};
  \draw[panel] (1.55,2.42) rectangle (16.05,5.60);
  \node[ptitle, anchor=east] at (15.78,5.30) {one intent; mode chosen per request};

  \node[chip, minimum width=1.70cm, minimum height=0.85cm] (intent) at (2.78,4.00)
    {\textbf{Intent}\\ \textbf{+ Scope}};

  \node[modetag, fill=slate, minimum width=1.45cm, minimum height=0.50cm] (mdef) at (4.88,4.72) {\textit{Default}};
  \node[chip, minimum width=1.85cm, minimum height=0.52cm] (inspect) at (6.78,4.72) {Inspect\\ Capabilities};
  \node[chip, minimum width=0.95cm, minimum height=0.52cm] (plan) at (8.33,4.72) {Plan};
  \node[chip, minimum width=1.15cm, minimum height=0.52cm] (exec) at (9.68,4.72) {Execute};
  \node[chip, minimum width=1.05cm, minimum height=0.52cm] (verify) at (11.03,4.72) {Verify};
  \node[chip, minimum width=1.20cm, minimum height=0.52cm] (hydrate) at (12.43,4.72) {Hydrate\\ Fields};

  \node[modetag, fill=amber, minimum width=1.45cm, minimum height=0.50cm] (mdeep) at (4.88,3.40) {\textit{Deep Search}};
  \node[chip, minimum width=1.85cm, minimum height=0.52cm] (sret) at (6.98,3.40) {Semantic\\ Retrieve};
  \node[chip, minimum width=1.85cm, minimum height=0.52cm] (eval) at (9.18,3.40) {Evaluate\\ Candidates};
  \node[chip, minimum width=1.85cm, minimum height=0.52cm] (accref) at (11.38,3.40) {Accept\\ or Refine};
  \draw[loop] (accref.south) to[out=-155,in=-25,looseness=0.40]
    node[below=0.5pt,note]{refine within bounds; session persists across turns} (sret.south);

  \node[outchip, minimum width=2.15cm, minimum height=1.05cm] (outA) at (14.72,4.00)
    {\textbf{Typed Results}\\ scenes, rows,\\ session state};

  \draw[tinyflow] (intent.east) -- (mdef.west);
  \draw[tinyflow] (intent.east) -- (mdeep.west);
  \draw[tinyflow] (mdef) -- (inspect);
  \draw[tinyflow] (inspect) -- (plan);
  \draw[tinyflow] (plan) -- (exec);
  \draw[tinyflow] (exec) -- (verify);
  \draw[tinyflow] (verify) -- (hydrate);
  \draw[tinyflow] (mdeep) -- (sret);
  \draw[tinyflow] (sret) -- (eval);
  \draw[tinyflow] (eval) -- (accref);
  \draw[tinyflow] (hydrate.east) -- (outA.170);
  \draw[tinyflow] (accref.east) -- (outA.190);

  \node[lane, anchor=center] at (0.68,1.32) {DIRECT\\ACCESS};
  \draw[panel] (1.55,0.42) rectangle (16.05,2.24);
  \node[ptitle, anchor=east] at (15.78,2.06) {one explicit contract per call};

  \node[chip, minimum width=2.70cm, minimum height=0.62cm] (sem) at (4.60,1.55)
    {\textbf{\textit{Semantic Search}}\\ meaning-based ranking};
  \node[chip, minimum width=2.70cm, minimum height=0.62cm] (qry) at (8.80,1.55)
    {\textbf{\textit{Query}}\\ typed predicates};
  \node[chip, minimum width=2.70cm, minimum height=0.62cm] (agg) at (13.00,1.55)
    {\textbf{\textit{Aggregate}}\\ grouped analysis};
  \node[outchip, minimum width=2.30cm, minimum height=0.40cm] (semo) at (4.60,0.72) {Ranked Scenes};
  \node[outchip, minimum width=2.30cm, minimum height=0.40cm] (qryo) at (8.80,0.72) {Scenes Matching Predicates};
  \node[outchip, minimum width=2.30cm, minimum height=0.40cm] (aggo) at (13.00,0.72) {Analytical Rows};
  \draw[tinyflow] (sem.south) -- (semo.north);
  \draw[tinyflow] (qry.south) -- (qryo.north);
  \draw[tinyflow] (agg.south) -- (aggo.north);

  \node[lane, anchor=center] at (0.68,-0.70) {\textit{ASK}};
  \draw[panel] (1.55,-1.30) rectangle (16.05,-0.10);
  \node[ptitle, anchor=east] at (15.78,-0.28) {answer built over retrieved, hydrated context};

  \node[chip, minimum width=2.00cm, minimum height=0.62cm] (aret) at (3.10,-0.76)
    {\textbf{Retrieve}\\ \textit{Search} or \textit{Deep Search}};
  \node[chip, minimum width=2.00cm, minimum height=0.62cm] (ahyd) at (5.75,-0.76)
    {\textbf{Hydrate}\\ timestamped context};
  \node[chip, minimum width=1.55cm, minimum height=0.62cm] (agen) at (8.18,-0.76)
    {\textbf{Generate}};
  \node[chip, minimum width=1.95cm, minimum height=0.62cm] (aval) at (10.45,-0.76)
    {\textbf{Validate}\\ references};
  \node[outchip, minimum width=2.15cm, minimum height=0.62cm] (outC) at (14.72,-0.76)
    {\textbf{Answer with}\\ \textbf{Source Shots}};
  \draw[tinyflow] (aret) -- (ahyd);
  \draw[tinyflow] (ahyd) -- (agen);
  \draw[tinyflow] (agen) -- (aval);
  \draw[tinyflow] (aval) -- (outC);
\end{tikzpicture}
\caption{The search surface as staged flows. High-level \emph{Search} takes
one intent and runs it through the mode chosen for the request: default mode
plans over declared capabilities, executes, optionally verifies with a
precision-oriented filter, and hydrates stored fields; \emph{Deep Search}
mode iterates semantic retrieval within configured bounds and persists the session
across turns. Direct access binds one explicit contract per call, each with
its own typed output. \emph{Ask} builds a validated answer over retrieved,
hydrated context and returns the shots that ground it.
Table~\ref{tab:contracts} gives the corresponding contract mapping.}
\label{fig:surface}
\end{figure}
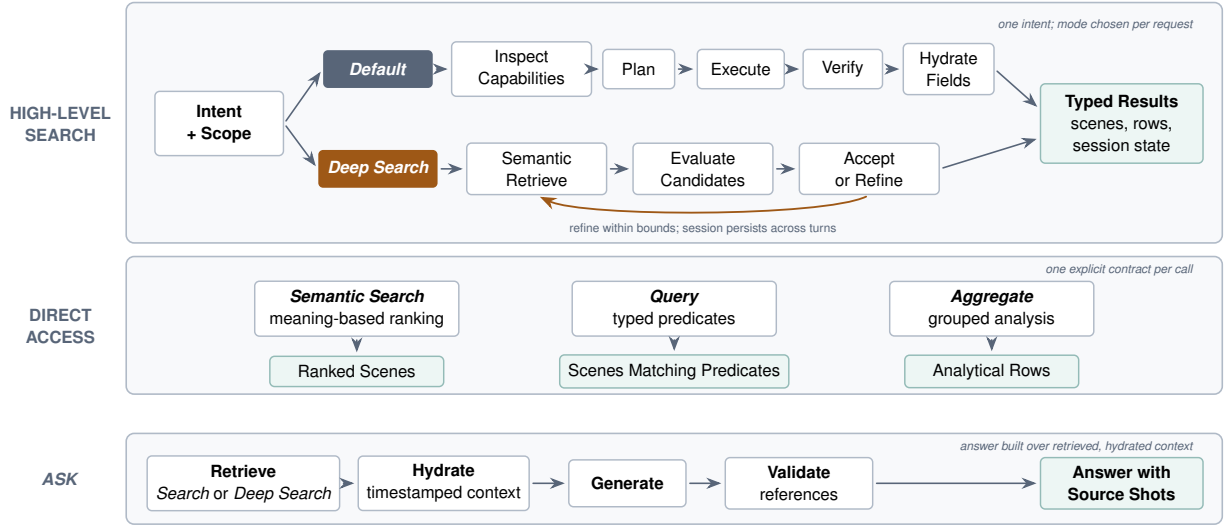

A single search operation cannot serve every information need without
collapsing distinctions that matter: planned retrieval, semantic similarity,
exact selection, grouped analysis, iterative investigation, and grounded
synthesis have different contracts. Related temporal-grounding work likewise
distinguishes moment retrieval, highlight detection, and query-conditioned
summarization as related but distinct tasks \citep{lin2023univtg}. The reference implementation therefore
exposes one high-level \emph{Search} with two modes, three direct-access
contracts, and \emph{Ask} (Figure~\ref{fig:surface}). The contracts are
complementary rather than successive quality levels; the right choice depends
on what the request needs to be true of its result. Table~\ref{tab:contracts}
states the mapping. Throughout the report, named contracts and modes are
italicized, while quoted italics, as in \qtext{label contains `cell phone'},
mark literal request text. The contract chosen for a request determines what enters
an agent's context: an analytical question should not return video, a
verification task should not return only a count, and a synthesis should keep
the scenes that ground it.

\begin{table}[H]
\centering
\small
\renewcommand{\arraystretch}{1.12}
\begin{tabular}{@{}p{0.235\textwidth}p{0.20\textwidth}p{0.305\textwidth}p{0.155\textwidth}@{}}
\toprule
\textbf{Information need} & \textbf{Contract} & \textbf{Returned context} & \textbf{Reference API} \\
\midrule
Meaning-based discovery & Semantic retrieval & Ranked, source-linked scenes & \emph{Semantic Search} \\
Exact conditions & Structured query & Scenes matching typed predicates & \emph{Query} \\
Counts and comparisons & Aggregation & Analytical rows, not intervals & \emph{Aggregate} \\
Planned retrieval & Orchestrated search & Merged, verified, hydrated scenes & \emph{Search} (default) \\
Iterative investigation & Stateful refinement & Results plus persisted session state & \emph{Search} (\emph{Deep Search}) \\
Grounded synthesis & Grounded answering & Answer plus its supporting scenes & \emph{Ask} \\
\bottomrule
\end{tabular}
\caption{The search surface as a mapping from information needs to typed
contracts. Reference-implementation names appear only to show one
realization; the contracts, not the API names, are the claim.}
\label{tab:contracts}
\end{table}

\paragraph{High-level \emph{Search}}
\emph{Search} accepts intent and scope rather than an explicit index
program. In \emph{default mode} the system inspects available indexes, their
schemas, and declared capabilities (P2); reasons jointly over the request and
that catalog to select relevant indexes and direct methods; constructs and
validates an execution plan; coordinates retrieval;
merges compatible results in source time (P4); and hydrates requested stored
fields into the returned shots \citep{videodb2026docs}. The planner cannot
assume an unavailable field
or operation. Default search may invoke a verifier after retrieval: a
precision-oriented filter that compares candidate shots with the original
intent and removes nonmatching candidates. It cannot recover evidence omitted by candidate retrieval, an asymmetry that matters for how the loop in
\S\ref{sec:loops} is designed.

\paragraph{\emph{Deep Search} mode}
In \emph{Deep Search} mode the system runs an agentic loop with verification
and improvement around semantic retrieval \citep{videodb2026deepsearch}: it evaluates candidate acceptance,
refines the retrieval strategy,
repeats retrieval within configured limits, and persists session state so an
investigation can continue across turns without discarding what has been
learned. \emph{Deep Search} currently reaches memory through its semantic-retrieval
bridge only; it does not execute \emph{Query} or \emph{Aggregate}. Within
that bridge, \emph{Deep Search} can combine semantic retrieval with supported metadata
filters derived from the request, then verify and improve the resulting
candidates across iterations. Structured and analytical operations remain
available through default \emph{Search} and direct access.

\paragraph{Direct contracts}
\emph{Semantic Search} is the explicit contract for meaning-based moment
retrieval: the caller supplies text and may target one or more semantic indexes
by name or ID. If neither index names nor index IDs are specified, VideoDB
searches all indexes in scope that declare semantic capability. The caller may
constrain eligibility with exact filters over fields those indexes expose;
results are ranked shots that preserve source identity and intervals.
\emph{Query} executes typed conditions over a queryable index when the caller
already knows field and operator semantics: exact categories, numeric thresholds, identifiers, conjunctions of declared predicates; matching temporal
records return as shots, so exact selection still ends in source-linked
evidence. \emph{Aggregate} answers analytical questions (counts, groups, facets, metrics) and returns \emph{rows}, not intervals. Keeping this type
distinct prevents a grouped count from being mistaken for playable evidence;
an application may use an aggregate to guide a later temporal request, but the
row itself is not a shot. The typed contracts are literal where-clauses over
declared fields: a \emph{Query} request against an object layer of the form
\qtext{label contains `cell phone', sorted by detection confidence} returns
exactly the matching shots, and an \emph{Aggregate} request of the form
\qtext{detection confidence at least 0.8, grouped by label, counted} returns
one row per label. Declared
operators (equality, containment, membership, numeric bounds, existence)
compose with AND, OR, and NOT over any field an index exposes
\citep{videodb2026docs}. Neither request invokes a model at query time; both
run entirely over stored understanding.

We also measured these direct stored-data contracts over five hours of indexed
video. Across 100 measured calls per contract, median latency was 0.443~s for
\emph{Query} and 0.380~s for \emph{Aggregate}. The TwelveLabs interface
used in \S\ref{sec:eval} does not expose corresponding direct contracts.
Appendix~\ref{app:operational} provides request examples and the measurement
setup.

\paragraph{\emph{Ask}}
\emph{Ask} builds a synthesized response from retrieved visual memory: it
retrieves through \emph{Search} or \emph{Deep Search}, hydrates timestamped artifact
context, generates an answer over that bounded context, validates references to
selected source entries, and optionally returns the corresponding shots. The
answer and the evidence remain distinct typed outputs: text serves synthesis;
shots remain inspectable source moments
\citep{videodb2026docs,gao2023enabling}.

\subsection{Indexed scenes and resolved scenes}\label{sec:resolved}

The scene spaces of \S\ref{sec:model} are built \emph{before} any question is
asked; queries arrive later and rarely respect stored boundaries. Temporal
grounding research likewise evaluates query-specific interval localization
\citep{soldan2022mad,lin2023univtg}. This forces a distinction that retrieval
systems usually leave implicit.

\begin{definition}[Indexed scene]\label{def:indexed}
An indexed scene is an analyzer-defined temporal unit created to support
understanding and retrieval.
\end{definition}

\begin{definition}[Resolved scene]\label{def:resolvedscene}
A resolved scene is a query-specific evidence interval produced after
reasoning over retrieved candidates.
\end{definition}

Externally, the system retrieves candidate scenes from one or more layers, can
inspect any relevant frame of the underlying media, and refines the final
temporal boundary of what it returns. The granularity used to understand and
index visual data therefore does not need to be the granularity returned as
evidence: a question about a three-second gesture can be answered from
half-minute indexed scenes, and a question about a full procedure can fuse many
short ones (Figure~\ref{fig:scenes}). A resolved scene is an output, not an
obligation: it need not be persisted, though a caller may store it as a new
artifact if it has lasting value. And because the delivery path reads the
format rather than stored segments (P7), a resolved scene is watchable the
moment it is resolved: its query-specific boundary is served as-is by the
streaming engine (\S\ref{sec:evidence}), never rounded to a stored unit.

\thesisbox{Search does not merely retrieve a previously stored scene. It can
construct the scene appropriate to the question, and play it the moment it
is constructed.}

\section{Closed-Loop Visual Search}\label{sec:loops}

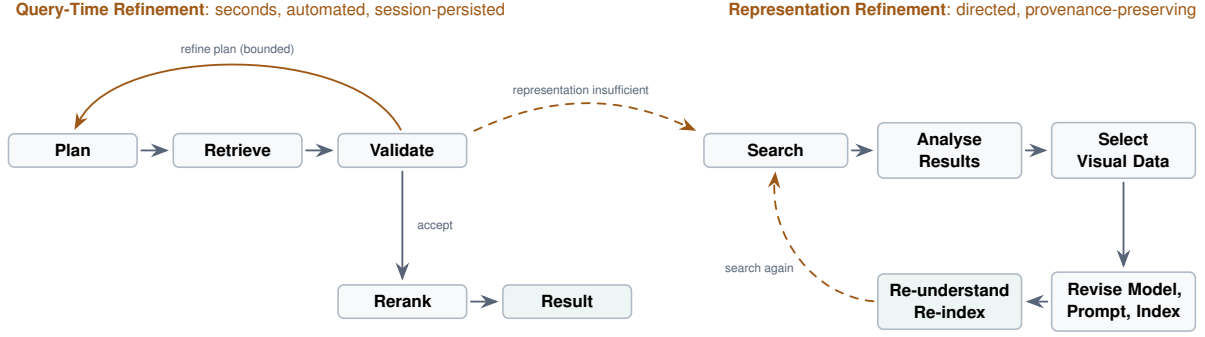
\begin{figure}[t]
\centering
\begin{tikzpicture}
  \node[note, text=amber, anchor=west, font=\sffamily\scriptsize] at (-0.88,3.00)
    {\textbf{Query-Time Refinement}: seconds, automated, session-persisted};
  \node[card, text width=1.42cm] (plan) at (0,1.15) {\textbf{Plan}};
  \node[card, text width=1.42cm] (retr) at (2.18,1.15) {\textbf{Retrieve}};
  \node[card, text width=1.42cm] (val) at (4.36,1.15) {\textbf{Validate}};
  \node[card, text width=1.42cm] (rer) at (4.36,-0.85) {\textbf{Rerank}};
  \node[cardT, text width=1.42cm] (res) at (6.54,-0.85) {\textbf{Result}};
  \draw[flow] (plan) -- (retr);
  \draw[flow] (retr) -- (val);
  \draw[flow] (val) -- node[right=2pt,note]{accept} (rer);
  \draw[flow] (rer) -- (res);
  \draw[loop] (val.north) to[out=118,in=62,looseness=0.78]
    node[above=0.5pt,note]{refine plan (bounded)} (plan.north);

  \node[note, text=amber, anchor=west, font=\sffamily\scriptsize] at (8.55,3.00)
    {\textbf{Representation Refinement}: directed, provenance-preserving};
  \node[card, text width=1.62cm] (srch2) at (9.30,1.15) {\textbf{Search}};
  \node[card, text width=1.62cm] (ana) at (11.60,1.15) {\textbf{Analyse}\\ \textbf{Results}};
  \node[card, text width=1.62cm] (sel) at (13.90,1.15) {\textbf{Select}\\ \textbf{Visual Data}};
  \node[card, text width=1.62cm] (rev) at (13.90,-0.85) {\textbf{Revise Model,}\\ \textbf{Prompt, Index}};
  \node[cardT, text width=1.62cm] (reidx) at (11.60,-0.85) {\textbf{Re-understand}\\ \textbf{Re-index}};
  \draw[flow] (srch2) -- (ana);
  \draw[flow] (ana) -- (sel);
  \draw[flow] (sel) -- (rev);
  \draw[flow] (rev) -- (reidx);
  \draw[loopd] (reidx.west) to[out=175,in=-95]
    node[below left=-2pt,note]{search again} (srch2.south);

  \draw[loopd] ($(val.east)+(0.05,0.24)$) to[out=25,in=155]
    node[above=-1pt,note,pos=0.48]{representation insufficient} ($(srch2.west)+(-0.05,0.24)$);
\end{tikzpicture}
\caption{The two loops of closed-loop visual search. \emph{Left:} in
query-time refinement, the system plans, retrieves, and validates, then
reranks into a result or refines the plan within configured bounds; session
state persists across turns (\emph{Deep Search}). \emph{Right:} in representation
refinement, analysis of search results leads a developer or an agent to
select visual data; change the analyzer, prompt, scene strategy, computed
property, or index; re-understand or re-index; and search again. The first loop is automated and demonstrated; the
second is enabled by the separation of understanding, memory, and access
(P1), and is not autonomous in the current system.}
\label{fig:twoloops}
\end{figure}

Search reads visual memory. It also \emph{tests} visual memory: a query whose
results are wrong or empty is information about the representations through which memory is being accessed: the analyzer, its prompt or sampling, the
scene strategy, the computed properties, the index. We call the resulting
discipline \emph{closed-loop visual search} and separate it into two loops with
different timescales, actors, and guarantees (Figure~\ref{fig:twoloops}).
Interactive video-database work provides a related precedent by synthesizing
compositional event queries from limited user feedback
\citep{zhang2023equivocal}.

\subsection{Query-time refinement}

The inner loop operates within a single investigation, and three operators
describe it exactly. Retrieval first draws candidates from the index layers
selected for the intent,
\[
C_0 \;=\; \bigcup_{k} \mathrm{Retrieve}(q,\, I_k),
\]
where $I_k$ ranges over the selected layers. A reasoning step then validates
relevance, rejects false positives, inspects supporting frames, and refines
the query within configured bounds,
\[
q_{t+1} \;=\; \mathrm{Refine}(q_t,\, C_t), \qquad
C_{t+1} \;=\; \mathrm{Retrieve}(q_{t+1},\, \mathcal{I}), \qquad t < T,
\]
over the available layers $\mathcal{I}$, and finally resolves the scene or
set of scenes appropriate to the question,
\[
\hat{S} \;=\; \mathrm{Resolve}(q,\, C_0, \ldots, C_T).
\]
$\mathrm{Resolve}$ is where Definition~\ref{def:resolvedscene} becomes
operational: candidates arrive from layers with non-aligned boundaries, and
the returned interval is constructed for the query rather than inherited from
ingestion (Figure~\ref{fig:resolve}). In process terms the loop plans,
retrieves, validates, then reranks into a result or refines the plan; it runs
within configured limits and persists as session state.

\begin{figure}[t]
\centering
\begin{tikzpicture}
  \foreach \y/\n in {2.30/Layer A, 1.70/Layer B, 1.10/Layer C, 0.10/Resolved}
    { \node[lane, anchor=east] at (0.55,\y) {\n};
      \draw[line width=0.5pt, draw=fog!55] (0.80,\y) -- (8.60,\y); }
  \draw[fill=rowa!22, draw=rowa, line width=0.6pt, rounded corners=1.5pt] (1.55,2.13) rectangle (3.75,2.47);
  \draw[fill=rowb!20, draw=rowb, line width=0.6pt, rounded corners=1.5pt] (2.45,1.53) rectangle (4.90,1.87);
  \draw[fill=rowd!18, draw=rowd, line width=0.6pt, rounded corners=1.5pt] (3.45,0.93) rectangle (5.70,1.27);
  \draw[dashed, draw=amber, line width=0.8pt] (3.05,2.72) -- (3.05,-0.34);
  \node[note, text=amber] at (3.05,-0.56) {refined boundary};
  \draw[fill=teal!22, draw=teal, line width=0.8pt, rounded corners=1.5pt] (3.05,-0.07) rectangle (5.25,0.27);
  \node[draw=none, fill=ink, text=white, rounded corners=2.5pt, align=center,
        font=\sffamily\scriptsize, inner sep=5pt] (reason) at (10.55,1.70)
    {\textbf{Reason}\\ inspect frames; refine query};
  \node[cardT, text width=2.55cm] (evlabel) at (10.55,0.10)
    {\textbf{Evidence Scene}\\ query-specific boundary};
  \draw[flow] (3.80,2.30) -- (reason.165);
  \draw[flow] (4.95,1.70) -- (reason.west);
  \draw[flow] (5.75,1.10) -- (reason.195);
  \draw[flow] (reason.south) -- (evlabel.north);
  \draw[line width=0.55pt, draw=fog] (evlabel.west) -- (5.35,0.10);
\end{tikzpicture}
\caption{From retrieved candidates to resolved evidence. Candidates arrive
from understanding layers with non-aligned boundaries; the reasoning step
inspects relevant frames and refines the query within bounds; the returned
evidence scene carries a query-specific boundary (dashed) that need not
coincide with any stored scene
(Definitions~\ref{def:indexed} and~\ref{def:resolvedscene}).}
\label{fig:resolve}
\end{figure}
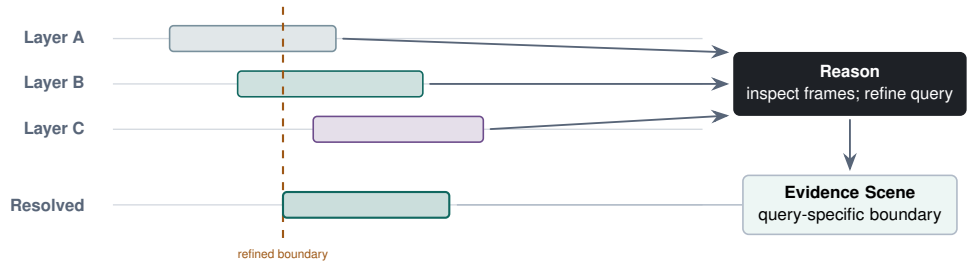 This is the
loop \emph{Deep Search} mode implements, and the open-source \emph{Deep Search} system built on
VideoDB makes it concrete and inspectable \citep{videodb2026deepsearch}. There,
a planner decomposes intent into subqueries, each targeting one or two named
understanding layers (in the reference indexing pipeline: location, action,
scene description, object description, topic, subplot summary, and
whole-source summary layers, plus structured facets such as detected objects,
emotion, and shot type); retrieval fans out with paraphrase variants and fuses
candidates per subquery; a join plan combines subqueries in source time; a
validator issues per-candidate verdicts (pass, ambiguous, fail) against the
original intent and, when nothing passes, emits structured feedback with a
bounded list of suggested plan operations; empty results trigger analysis and
constraint relaxation along a declared fallback order rather than silent
failure; an interpreter turns user steering (free text, choices on clarifying questions, \qtext{more like this} examples) into typed deltas on the plan; and the
session, including plan history and verdicts, persists across turns. The loop
composes retrieval, critique, and revision in the spirit of reflective
retrieval-augmented systems \citep{asai2024selfrag,yao2023react}, but its moves
are typed plan edits over declared capabilities rather than free-form tool
calls. Because the verifier is precision-oriented (\S\ref{sec:search}), the
loop's recovery mechanism for \emph{missing} evidence is plan revision (new subqueries, relaxed constraints, different layers), not post-hoc filtering.

\subsection{Representation refinement}

The outer loop operates on memory itself:

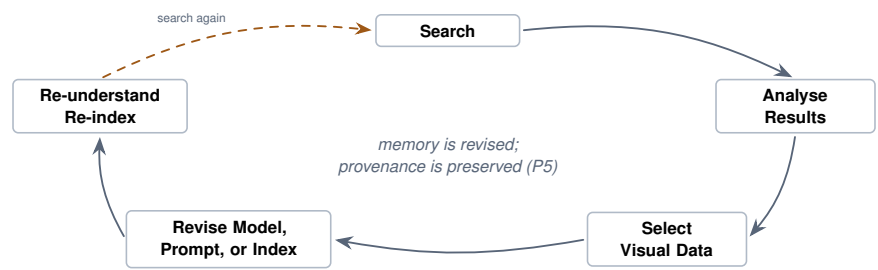
\begin{center}
\begin{tikzpicture}[
  ringchip/.style={draw=fog, line width=0.6pt, rounded corners=2pt, fill=white,
    align=center, font=\sffamily\scriptsize, inner sep=3.6pt}]
  \node[ringchip, minimum width=1.9cm] (r1) at (0,1.52) {\textbf{Search}};
  \node[ringchip, text width=1.85cm] (r2) at (4.60,0.52) {\textbf{Analyse}\\ \textbf{Results}};
  \node[ringchip, text width=1.85cm] (r3) at (2.90,-1.24) {\textbf{Select}\\ \textbf{Visual Data}};
  \node[ringchip, text width=2.45cm] (r4) at (-2.90,-1.24) {\textbf{Revise Model,}\\ \textbf{Prompt, or Index}};
  \node[ringchip, text width=2.05cm] (r5) at (-4.60,0.52) {\textbf{Re-understand}\\ \textbf{Re-index}};
  \node[font=\sffamily\scriptsize\itshape, text=slate, align=center] at (0,-0.15)
    {memory is revised;\\ provenance is preserved (P5)};
  \draw[flow] (r1.east) to[bend left=16] (r2.north);
  \draw[flow] (r2.south) to[bend left=16] (r3.east);
  \draw[flow] (r3.west) to[bend left=10] (r4.east);
  \draw[flow] (r4.west) to[bend left=16] (r5.south);
  \draw[loopd] (r5.north) to[bend left=16]
    node[above left=-1pt, note]{search again} (r1.west);
\end{tikzpicture}
\end{center}
A developer (or an agent workflow acting under a developer's direction) observes that a class of queries fails; localizes the failure to a
representation (a prompt that never mentions dock doors; sampling too sparse to
catch a gesture; a missing computed property; an absent domain analyzer);
selects the affected sources or intervals; applies a different analyzer
configuration or derives a new index from artifacts that already exist (P1);
and re-runs the search. VOCAL-UDF demonstrates a related, more automated path
in which a video-query system identifies missing analytic concepts and
constructs or selects new UDFs \citep{zhang2025vocaludf}. Provenance (P5) makes
the change inspectable: old and new views coexist, can be compared on the same
queries, and can be retired explicitly. This loop is \emph{enabled} by the
infrastructure; it is why understanding, memory, and access are kept separate. But we do not claim it is autonomous. In the current system, accepted results, rejected entries, edits,
and downstream outcomes do not yet become durable learning signals that update
future understanding and retrieval on their own (\S\ref{sec:limits}).

Search is therefore both the read interface to visual memory and the feedback
mechanism for improving the representations through which memory is accessed.
The two loops share one currency: evidence. The inner loop decides whether
retrieved intervals answer the question; the outer loop decides whether the
memory could have contained the answer at all.

\section{Model or System: Two Ways to Build Video Search}\label{sec:architectures}

Two design philosophies currently compete to define video search, and they
disagree about where quality comes from.

\paragraph{The video-native model}
The first philosophy treats video as a first-class perceptual signal and
trains an end-to-end foundation model for it. A representative commercial
example is Twelve Labs: Marengo, a multimodal embedding model that maps video,
audio, image, and text into one unified embedding space and powers its search
and embedding APIs, and Pegasus, a video-first language model for generation
\citep{twelvelabs2025marengo}. The interface this philosophy produces is an API
around a trained artifact: as publicly documented, an application creates an index by naming a
model version and enabling modalities, and the rest (how the timeline is segmented, how frames are sampled, how content is embedded, how results are ranked) is internal to the model and fixed at training time
\citep{twelvelabs2026docs}. The representation is the product; improving the
system means training the next model, and adopting it means re-indexing
through the API.

\paragraph{The visual data infrastructure}
The second philosophy, the one this report develops, treats search quality
as a property of the pipeline rather than of any single model. Every decision
the video-native model internalizes is exposed here as a system choice:
segmentation policy and scene granularity; frame rate and sampling per
analyzer; which model produces each understanding layer, at which capability
and cost tier; the prompt and output schema of each analyzer; the embedding
model behind each semantic index; the ranker and fusion policy at retrieval;
which computed properties and domain analyzers exist at all. The stored
understanding is itself visible: artifacts are data the user can read,
export, and re-index their own way (P1--P3). And because nothing in the data
model assumes a finished source (P6), live streams are first-class: scenes
append while the event is still happening, and search returns playable evidence of moments ago, a capability the uploaded-asset API model does not express. In the open-source \emph{Deep Search} pipeline, these degrees of freedom are
concrete configuration: per-stage model routing, per-layer index construction,
detector on or off, prompts per enrichment stage \citep{videodb2026deepsearch}.

Table~\ref{tab:architectures} summarizes the contrast. It should be read as a
comparison of documented interfaces, not as an independent assessment of model
quality. The two philosophies also compose rather than exclude one another. From the
infrastructure's point of view a video-native model is simply one more analyzer in the portfolio; and this is not hypothetical: Twelve Labs' models
run as a selectable analyzer inside VideoDB's real-time pipeline, a pairing
both companies document \citep{twelvelabs2025videodb}. The architectural
question is therefore not whether video-native models are good. It is where
the remaining degrees of freedom should live: frozen inside a trained
artifact, or exposed by the system that operates it.

\begin{table}[t]
\centering
\footnotesize
\renewcommand{\arraystretch}{1.16}
\begin{tabular}{@{}p{0.215\textwidth}p{0.345\textwidth}p{0.360\textwidth}@{}}
\toprule
 & \textbf{Video-native model API} & \textbf{Visual data infrastructure} \\
 & (Twelve Labs, as documented) & (VideoDB, this report) \\
\midrule
Primary artifact & A trained model behind an API & A logical data format (VDB) and a pipeline over many models \\
Understanding produced by & One proprietary model family (Marengo embeddings; Pegasus generation) & An open portfolio: ASR, detectors, OCR, VLMs of any tier, domain models; video-native models included \\
Segmentation \& sampling & Internal to the model & Chosen per analyzer: shot- or time-based scenes, frame rate, frames per scene \\
Embedding space & Unified, fixed at training time; exportable to an external vector stack via a separate embed API & Chosen per semantic index; replaceable without redoing understanding \\
Prompts \& output schemas & Not exposed at indexing & User-defined per analyzer; open schemas \\
Ranking & Internal & Selectable ranker, reranker, and fusion policy \\
Visibility of understanding & Embeddings and internal state opaque & Artifacts readable and exportable; users index their own way \\
New access path over old media & Re-index through the model & Derive a new index from stored artifacts; no re-analysis \\
Live streams & Managed search documented over uploaded, indexed assets; real-time workflows documented through the VideoDB integration & First-class sources; scenes append in real time; evidence from moments ago \\
Evidence output & Timestamped segments with playback metadata for indexed sources; cross-source composition left to the application & Query-resolved scenes; programmable composition (sequence, stack, overlay) into one playable evidence stream, archived or live \\
Improvement path & Train and migrate to the next model version & Recompose: swap an analyzer or index; provenance keeps old and new comparable \\
\bottomrule
\end{tabular}
\caption{Two architectures for search over video. The left column reflects the
public product interface of a commercial video-native system as of July 2026
\citep{twelvelabs2025marengo,twelvelabs2026docs}; the right column is the
infrastructure developed in this report. The columns compose: a video-native
model can serve as one analyzer inside the infrastructure
\citep{twelvelabs2025videodb}.}
\label{tab:architectures}
\end{table}

\subsection{Why search is a systems problem today}\label{sec:tokenization}

There is a structural reason to treat video search as a systems problem. Text
models consume discrete, one-dimensional token sequences, although tokenizers
differ across model families. Visual models must additionally discretize two
spatial axes and time, and current systems make different choices. Image models
patch space \citep{dosovitskiy2021image}; video models extend patches across
time into spatio-temporal tokens \citep{arnab2021vivit}; embedding systems
compress segments into one or several vectors
\citep{bain2021frozen,ni2022xclip,twelvelabs2025marengo}; and long-video models
adaptively reduce spatial and temporal tokens before reasoning
\citep{shen2025longvu}. Each video-native model commits to an internal
discretization and exposes only part of that choice through its interface. When that
frozen choice is wrong for a workload (too coarse for a three-second gesture, too short-horizon for a forty-minute procedure, blind to a domain's structures), the application cannot fix it without a different model.

While temporal-visual representations remain workload-dependent, retrieval
quality over the visual world can also be influenced by decisions a system can
expose and revise (segmentation, sampling, model portfolio, index composition, ranking) and by the loops that revise them against evidence
(\S\ref{sec:loops}). Search engine design over video is, today, a systems
problem rather than a training-time optimization.

We take the image to be the
native unit of stored visual information: a video is images ordered in time,
joined by sound. The infrastructure is built on that position: understanding is anchored to frames and intervals on the source-time axis, and temporal
structure is reconstructed by alignment across layers rather than baked
irreversibly into any single representation.

\section{Empirical Study}\label{sec:eval}

Section~\ref{sec:architectures} locates specialization in two different places:
a video-native system concentrates it inside a trained representation, whereas
visual data infrastructure exposes segmentation, sampling, schemas, indexes,
and ranking as choices that can be composed around stored understanding. That
architectural contrast leaves an empirical question. Can a pipeline assembled
from general-purpose components operate in the same retrieval-quality regime as
a managed video-native search path?

We answer with a shared-corpus comparison between VideoDB and TwelveLabs. The
unit of comparison is the retrieval path an application invokes, not an
embedding model removed from its surrounding system. VideoDB uses
collection-level semantic retrieval followed by reranking. TwelveLabs uses its
managed Marengo~3.0 search path
\citep{twelvelabs2025marengo,twelvelabs2026search}. Each path is configured
through the controls its interface provides. The comparison evaluates the
complete semantic-retrieval path exposed by each system. VideoDB's other search
contracts are outside this experiment.

\subsection{Corpus and Evaluation Design}

The evaluation uses a shared corpus. Both systems indexed the same source
videos and, for each request, received the same query text and dataset-defined
relevant item. To keep the comparison from resting on one content regime, the
corpus combines four public datasets and 9{,}800+ natural-language queries.
MSVD contributes short open-domain clips paired with paraphrastic descriptions
\citep{chen2011collecting}. YouCook2 contributes temporally localized steps in
instructional cooking videos \citep{zhou2018towards}. VATEX contributes diverse
web video with rich multilingual captions \citep{wang2019vatex}. MSR-VTT
contributes broad-category web clips paired with multiple natural-language
descriptions \citep{xu2016msrvtt}. The four regimes stress different retrieval
behavior: short semantic correspondence, procedural action, multi-action
description, and broad-category multimedia captioning.

\begin{table}[H]
\centering
\small
\begin{tabular}{@{}lrrr@{}}
\toprule
\textbf{Dataset} & \textbf{Videos} & \textbf{Hours} & \textbf{Queries} \\
\midrule
MSVD         & 213 &  7.3567 & 5{,}476 \\
YouCook2     & 149 & 12.2989 & 1{,}152 \\
VATEX        & 336 & 15.0000 & 3{,}000 \\
MSR-VTT      & 187 & 12.9808 &   206 \\
\midrule
\textbf{Total} & \textbf{885} & \textbf{47.6364} & \textbf{9{,}834} \\
\bottomrule
\end{tabular}
\caption{Exact composition of the evaluated corpus.}
\label{tab:benchmark-corpus}
\end{table}

\subsection{Evaluated Retrieval Paths}\label{sec:benchmark-config}

With corpus and relevance judgments fixed, the remaining choices concern how
each system represents, retrieves, and ranks the shared media.
Table~\ref{tab:benchmark-terms} summarizes those choices. To keep candidate
pools dataset-specific, VideoDB uses one collection per dataset, while
TwelveLabs uses one index per dataset. VideoDB makes scene construction,
artifact shape, index composition, and reranking configurable. Marengo keeps
the corresponding decisions provider-managed.

\begin{table}[H]
\centering
\small
\renewcommand{\arraystretch}{1.16}
\begin{tabular}{@{}p{0.105\textwidth}p{0.475\textwidth}p{0.33\textwidth}@{}}
\toprule
\textbf{Term} & \textbf{VideoDB configuration} & \textbf{Marengo 3.0 configuration} \\
\midrule
\term{scene} & Five-second source intervals & Provider-managed \\
\term{analyzer} & English timed transcripts plus \texttt{ultra}-tier VLM analysis of six sampled frames per scene with aligned transcript context & Provider-managed visual and audio analysis \\
\term{artifact} & Strict-JSON scene records with fields configured for each use case (Table~\ref{tab:dataset-indexes}) & Internal representation without prompt or schema controls \\
\term{index} & One collection per dataset, with a separate semantic index for each VLM-generated artifact field & One index per dataset; representation and index internals provider-managed \\
\term{search} & The unmodified caption retrieves 50 scene candidates through collection-level semantic search across the field-specific indexes in the dataset collection & The same caption retrieves 50 clip-grouped results from the dataset index using visual, audio, and transcript signals \\
\term{reranking} & BGE Gemma ranks candidates using concatenated scene fields & Provider-managed \\
\term{evidence} & Ranked, source-aligned shots with streamable evidence generated at a sub-second client-observed median & Ranked clips with timestamps \\
\bottomrule
\end{tabular}
\caption{Retrieval configurations compared in the shared-corpus experiment.}
\label{tab:benchmark-terms}
\end{table}

\paragraph{From source media to indexed representation}
VideoDB first constructs the representation that retrieval will search. Each
dataset occupies a separate collection, and every source in that collection is
divided into five-second scenes. The built-in
\texttt{spoken\_words} analyzer produces a timed transcript artifact and was
configured here for English \citep{videodb2026artifacts}. For each scene, a VLM
analyzer at the \texttt{ultra} tier receives six uniformly sampled frames
together with the aligned transcript and produces a strict-JSON
\texttt{scene} artifact.

The infrastructure does not prescribe a single domain schema. Artifact fields
can be configured for the information a use case needs
(Table~\ref{tab:dataset-indexes}). In this evaluation, MSVD represents compact
subject--action content, YouCook2 represents procedural state and ingredients,
VATEX represents richer ordered activity, and MSR-VTT represents broad-category
multimedia descriptions, including speech-oriented content. Each VLM-generated
artifact field is indexed separately, so
every dataset collection contains multiple semantic indexes. Collection-level
search spans those indexes. Appendix~\ref{app:benchmark-prompts} reproduces the exact
prompts and schemas. This makes adaptation an explicit prompt-and-schema choice
rather than requiring a new retrieval architecture or foundation model.

\begin{table}[H]
\centering
\small
\renewcommand{\arraystretch}{1.12}
\begin{tabular}{@{}p{0.16\textwidth}p{0.08\textwidth}p{0.67\textwidth}@{}}
\toprule
\textbf{Dataset} & \textbf{Fields} & \textbf{Semantically indexed representation} \\
\midrule
MSVD & 7 & Scene and short captions, action, location, characters, objects, and on-screen text \\
YouCook2 & 7 & Scene description, action, recipe step, ingredients, tools, food state, and on-screen text \\
VATEX & 8 & Scene and rich captions, ordered actions, location, characters, objects, on-screen text, and speech \\
MSR-VTT & 8 & Scene and short captions, action, location, characters, objects, on-screen text, and speech \\
\bottomrule
\end{tabular}
\caption{Use-case-specific VideoDB artifact schemas.}
\label{tab:dataset-indexes}
\end{table}

\paragraph{From candidates to ranked evidence}
Semantic retrieval determines which scenes enter the candidate set; reranking
determines their order. For this experiment, semantic retrieval uses
$\mathrm{top\_k}=50$, so each query supplies 50 candidates to the reranker.
This candidate depth is an evaluation setting, not a system limit. VideoDB
makes both the ranking model and the construction of candidate text
configurable. We evaluate four rerankers under two input modes: BGE Gemma
\citep{baai2024bgererankergemma}, mxbai Rerank Base v2 and Large v2
\citep{li2025prorank}, and mxbai Edge ColBERT v0 32M
\citep{takehi2025mxbai}. \emph{Match Exact} uses only the artifact field that
produced the semantic match.
\emph{Concatenate} combines all VLM text fields for the scene, allowing the
reranker to judge the candidate against the fuller stored representation.

Crossing the four rerankers with the two input modes yields the eight operating
points in Table~\ref{tab:reranker-operating-points}. The table reports
four-dataset macro recall alongside median component latency;
Appendix~\ref{app:reranker-study} reports every dataset-level result. BGE Gemma
with Concatenate is highest at each cutoff affected by candidate ordering:
73.09 at R@1, 83.39 at R@3, and 91.20 at R@10. All configurations have the same
96.07 R@50 because they reorder the same 50 candidates. We therefore fix BGE
Gemma with Concatenate for every VideoDB value in Table~\ref{tab:results}. Its
median component latency is 0.716~s, while the alternatives range from 0.289 to
1.288~s. The selected point favors retrieval quality rather than minimum
latency. Because reranking consumes fields already present in the artifact, a
deployment can change this operating point without regenerating the stored
understanding.

\begin{table}[H]
\centering
\footnotesize
\setlength{\tabcolsep}{3.8pt}
\begin{tabular}{@{}llccccr@{}}
\toprule
& & \multicolumn{4}{c}{\textbf{Macro recall (\%)}} & \\
\cmidrule(lr){3-6}
\textbf{Model} & \textbf{Input mode} & \textbf{R@1} & \textbf{R@3} & \textbf{R@10} & \textbf{R@50} & \textbf{Median (s)} \\
\midrule
BGE Gemma & Match Exact & 67.66 & 80.74 & 90.04 & 96.07 & 1.288 \\
\textbf{BGE Gemma} & \textbf{Concatenate} & \textbf{73.09} & \textbf{83.39} & \textbf{91.20} & 96.07 & 0.716 \\
mxbai Base v2 & Match Exact & 66.10 & 72.95 & 84.61 & 96.07 & 0.289 \\
mxbai Base v2 & Concatenate & 69.44 & 78.91 & 87.98 & 96.07 & 1.117 \\
mxbai Large v2 & Match Exact & 65.56 & 68.78 & 78.86 & 96.07 & 0.336 \\
mxbai Large v2 & Concatenate & 67.20 & 71.91 & 82.37 & 96.07 & 0.457 \\
mxbai Edge ColBERT & Match Exact & 65.06 & 77.49 & 88.71 & 96.07 & 0.407 \\
mxbai Edge ColBERT & Concatenate & 68.85 & 80.28 & 89.14 & 96.07 & 0.512 \\
\bottomrule
\end{tabular}
\caption{Macro recall and median component latency for the eight reranking
operating points. Recall is the unweighted mean across the four datasets. Bold
marks the selected configuration and its leading early-rank recall.}
\label{tab:reranker-operating-points}
\end{table}

\paragraph{The managed video-native path}
The benchmark uses one TwelveLabs index per dataset. The four indexes apply the
same provider-managed Marengo~3.0 configuration: visual and audio information
is indexed, then search operates over visual, audio, and transcript signals. Transcript search combines lexical and semantic
matching with OR. Results are grouped by clip, and the first 50 are returned
\citep{twelvelabs2026search}. Marengo does not expose dataset-specific prompts or artifact schemas, so the
same configuration remains in place across datasets
\citep{twelvelabs2026docs}. VideoDB makes use-case-specific representation a
configurable system choice; Marengo keeps it provider-managed.

\subsection{Retrieval Quality}

We measure retrieval quality with Recall@$k$, the percentage of queries for
which a relevant item appears among the first $k$ results. The macro-average is
the unweighted mean of the four dataset values. At query time, VideoDB applies
collection-level semantic retrieval followed by the fixed BGE Gemma
Concatenate reranking configuration. TwelveLabs applies the Marengo~3.0 path
above. We refer to the complete fixed VideoDB path simply as \emph{VideoDB}.
Table~\ref{tab:results} reports aggregate recall from the two runs.

\begin{table}[H]
\centering
\small
\setlength{\tabcolsep}{5.4pt}
\begin{tabular}{@{}lcccccccc@{}}
\toprule
 & \multicolumn{4}{c}{\textbf{VideoDB}} & \multicolumn{4}{c}{\textbf{TwelveLabs}} \\
\cmidrule(lr){2-5}\cmidrule(l){6-9}
\textbf{Dataset} & R@1 & R@3 & R@10 & R@50 & R@1 & R@3 & R@10 & R@50 \\
\midrule
MSVD         & \textbf{70.10} & \textbf{80.73} & 89.17 & 96.39 & 67.86 & 78.22 & \textbf{89.48} & \textbf{97.10} \\
YouCook2     & \textbf{65.96} & \textbf{80.98} & \textbf{93.48} & \textbf{97.87} & 47.07 & 65.56 & 84.18 & 96.41 \\
VATEX        & 83.46 & 90.28 & 95.24 & 97.77 & \textbf{85.43} & \textbf{92.40} & \textbf{97.30} & \textbf{99.43} \\
MSR-VTT      & \textbf{72.82} & \textbf{81.55} & \textbf{86.89} & 92.23 & 62.62 & 72.33 & 85.44 & \textbf{92.72} \\
\midrule
Macro-average & \textbf{73.09} & \textbf{83.39} & \textbf{91.20} & 96.07 & 65.75 & 77.13 & 89.10 & \textbf{96.42} \\
\bottomrule
\end{tabular}
\caption{Complete-system semantic-retrieval comparison using 9{,}800+ natural-language
queries. Values are percentages; the macro row is the unweighted arithmetic
mean of the four dataset rows. Bold marks the higher system for each dataset
and cutoff.}
\label{tab:results}
\end{table}

\paragraph{The observed pattern}
Table~\ref{tab:results} contains two results. First, on the unweighted
macro-average, VideoDB is higher through the first ten returned results:
73.09 versus 65.75 at R@1, 83.39 versus 77.13 at R@3, and 91.20 versus 89.10
at R@10. The margin narrows as the cutoff grows, and TwelveLabs is higher at
R@50, 96.42 versus 96.07.

Second, the ordering is dataset-dependent rather than universal. YouCook2 is
the clearest separation in VideoDB's favor: it is higher at every reported
cutoff, including by 18.89 points at R@1. MSVD divides by cutoff, with VideoDB
higher at R@1 and R@3 and TwelveLabs higher at R@10 and R@50. MSR-VTT follows
a similar early-rank pattern: VideoDB is higher through R@10, while TwelveLabs
is higher at R@50. TwelveLabs is higher at every reported cutoff on VATEX.

\paragraph{What the comparison establishes}
The macro-average result answers the empirical question posed at the beginning
of the section: a configurable pipeline of general-purpose components can
operate in the same quality regime as a managed video-native retrieval path and
can be higher at the early ranks. The dataset-level variation is equally
important. The result is not a universal ordering of the two systems; it shows
that retrieval quality depends on how a representation and its ranking path fit
the query distribution.

YouCook2 is consistent with the value of workload-specific representation: its
prompt, schema, semantic index, and reranking input are all shaped around
procedural cooking language. Because those choices move together, the study
does not assign the advantage to any one of them. The supported architectural
conclusion is compositional. Specialization can be assembled through exposed
choices in understanding, indexing, and ranking; it need not reside exclusively
inside one end-to-end video model. Conversely, a video-native model remains
compatible with the same infrastructure as a replaceable analyzer
(\S\ref{sec:architectures}).

\subsection{Configuring the Searchable Representation}\label{sec:representation-choices}

The complete-system comparison fixes one VideoDB retrieval path so that every
request encounters the same representation and ranking procedure. That fixed
path is an evaluation choice, not a constraint of the infrastructure. We next
vary two decisions made when the searchable representation is constructed: the
analyzer that writes each scene artifact, and the temporal and visual sampling
from which the artifact is written. In both studies, the dataset prompts and
schemas, collection and index topology, semantic-search depth, and BGE Gemma
Concatenate reranker remain fixed. Both studies use an evaluation subset drawn
from MSVD, YouCook2, and VATEX, comprising 15 hours of video and more than
4{,}000 natural-language queries. In this subsection, video Recall@$k$ counts a query
as a hit when at least one of the first $k$ returned scenes belongs to the
dataset-defined source video; the reported macro is the unweighted mean across
the three datasets.

\subsubsection{Analyzer Choice}

An analyzer determines what language is stored in each scene artifact and,
therefore, what the semantic indexes can retrieve. The complete-system path in
\S\ref{sec:benchmark-config} divides each source into five-second scenes,
samples six frames uniformly from each scene, and supplies the analyzer with
those frames and the aligned transcript. This study preserves that scene input and substitutes
one of six VLMs as the analyzer: Qwen 3.5 9B and 27B
\citep{qwen2026qwen359b,qwen2026qwen3527b}; GPT-5.4 Mini and GPT-5.5
\citep{openai2026gpt54mini,openai2026gpt55}; and Gemini 3.1 Pro and Gemini 3.6
Flash \citep{google2026gemini31pro,google2026gemini36flash}. Each VLM receives
the same frames, transcript context, dataset-specific prompt, and output schema; the
resulting artifacts then pass through the same index topology,
$\mathrm{top\_k}=50$ semantic retrieval, and BGE Gemma Concatenate reranker.
Table~\ref{tab:vlm-choice} reports the unweighted three-dataset macro video
recall.

\begin{table}[H]
\centering
\small
\setlength{\tabcolsep}{6.2pt}
\begin{tabular}{@{}lrrrr@{}}
\toprule
\textbf{Analyzer} & \textbf{R@1} & \textbf{R@3} & \textbf{R@10} & \textbf{R@50} \\
\midrule
Qwen 3.5 9B FP8       & 32.87 & 43.69 & 54.61 & 62.28 \\
Qwen 3.5 27B FP8      & 72.87 & 84.17 & 92.38 & 97.23 \\
GPT-5.4 Mini          & 77.44 & 86.96 & 94.11 & 97.53 \\
Gemini 3.1 Pro        & 79.52 & 88.17 & 94.91 & 98.15 \\
GPT-5.5               & 80.33 & 88.84 & 95.10 & 98.40 \\
Gemini 3.6 Flash      & \textbf{80.59} & \textbf{89.28} & \textbf{95.55} & \textbf{98.53} \\
\bottomrule
\end{tabular}
\caption{Macro video recall for six analyzer choices under a fixed downstream
retrieval path. Values are percentages and are the unweighted mean across the
three datasets. Bold marks the highest observed value at each cutoff.}
\label{tab:vlm-choice}
\end{table}

Table~\ref{tab:vlm-choice} contains two patterns. First, analyzer choice can
change whether the relevant source enters the candidate set at all. Qwen 3.5
9B reaches 62.28 at R@50, while each of the other analyzers reaches at least
97.23. BGE Gemma only reorders the 50 candidates produced for a given
representation, so it cannot recover a source that is absent from that set.
The R@50 gap therefore locates this difference before reranking, in the
artifacts and semantic candidate generation.

Second, the higher-performing analyzers occupy a much narrower quality
band. Across GPT-5.4 Mini, Gemini 3.1 Pro, GPT-5.5, and Gemini 3.6 Flash, macro
R@1 ranges from 77.44 to 80.59 and R@50 from 97.53 to 98.53. Gemini 3.6 Flash
has the highest observed macro at every cutoff, but its margin over GPT-5.5 is
at most 0.45 points through R@10 and 0.13 points at R@50. Differences of this
size do not establish a universal model ordering, particularly when the same
dataset-specific prompts need not be equally well matched to every model
family. The result makes the role of model-agnostic infrastructure precise:
the analyzer can be replaced without changing the surrounding retrieval path,
while the retrieval value of the artifacts it produces remains visible and
measurable.

\subsubsection{Scene Duration and Frames per Scene}

A scene artifact represents a bounded interval of source time using a finite
sample of frames. Scene duration controls how much time the artifact covers;
the frame count controls how densely that interval is shown to the analyzer.
To examine these choices, we fix GPT-5.5 and build nine searchable
representations by pairing scene durations of 2, 5, and 10 seconds with 2, 6,
and 10 uniformly sampled frames per scene. For every pairing, GPT-5.5 receives
the corresponding frames and aligned transcript, while the prompts, artifact
schemas, index topology, $\mathrm{top\_k}=50$ semantic retrieval, and BGE
Gemma Concatenate reranker remain unchanged. Table~\ref{tab:scene-frame-choice}
reports macro video recall for the nine representations.

\begin{table}[H]
\centering
\small
\setlength{\tabcolsep}{6.0pt}
\begin{tabular}{@{}rrrrrr@{}}
\toprule
\textbf{Duration (s)} & \textbf{Frames} & \textbf{R@1} & \textbf{R@3} & \textbf{R@10} & \textbf{R@50} \\
\midrule
 2 &  2 & 77.65 & 86.77 & 94.37 & 97.75 \\
 2 &  6 & 78.79 & 87.44 & 94.21 & 97.96 \\
 2 & 10 & 78.97 & 87.19 & 94.04 & 97.79 \\
 5 &  2 & 77.61 & 86.60 & 94.95 & 97.74 \\
 5 &  6 & \textbf{80.33} & 88.84 & 95.10 & \textbf{98.40} \\
 5 & 10 & 79.54 & 87.95 & 94.06 & 97.39 \\
10 &  2 & 77.65 & 86.93 & 94.11 & 97.44 \\
10 &  6 & 79.64 & 88.87 & \textbf{95.59} & 98.09 \\
10 & 10 & 79.05 & \textbf{89.26} & 94.58 & 97.65 \\
\bottomrule
\end{tabular}
\caption{Macro video recall across scene-duration and frames-per-scene choices
with GPT-5.5 fixed. Values are percentages and are the unweighted mean across
the three datasets. Bold marks the highest observed value at each cutoff.}
\label{tab:scene-frame-choice}
\end{table}

For the source-video retrieval queries represented by these three datasets,
the observed changes are modest. Across the nine configurations, macro recall
spans 2.72 points at R@1, 2.66 at R@3, 1.55 at R@10, and 1.01 at R@50. No
duration leads at every cutoff: five seconds and six frames is highest at R@1
and R@50, ten seconds and ten frames is highest at R@3, and ten seconds and six
frames is highest at R@10. Averaged over the three durations, six frames has
the highest observed recall at every cutoff, while increasing the sample to
ten frames provides no consistent gain on this subset.

We therefore use five-second scenes and six frames as a balanced operating
point for the complete-system comparison, not as a universal optimum. The
narrow spread in this study does not imply that scene construction is generally
inconsequential: duration and frame count still determine the temporal extent
and visual detail stored in each artifact, as well as the amount of indexing
work. Workloads with different event timescales or information requirements
may favor a different point.

\section{Future and a Research Agenda}\label{sec:limits}

\paragraph{Persistent, not adaptive}
The central limitation is that the infrastructure currently provides
\emph{persistent} visual memory, not memory that improves itself through use.
Artifacts and indexes survive requests, and \emph{Deep Search} persists investigation
state; but accepted results, rejected entries, edits, and downstream outcomes
do not yet become durable learning signals that update future understanding and
retrieval. The end state we envision is a \emph{self-improving} search
system: one that treats every investigation as a training episode and closes
the outer loop of \S\ref{sec:loops} autonomously. Reinforcement-learning
formulations fit that loop's structure naturally: retrieval plans, analyzer
and prompt selection, segmentation policies, and index composition are
actions; evidence accepted, corrected, or discarded downstream is the
grounded reward; and the provenance-preserving format supplies the replay
substrate such learning requires, since old and new representations coexist
and remain comparable on identical queries (P5). An adaptive system could collect explicit signals (relevance judgments, corrections, ratings) and implicit ones (inspected shots, playback duration, compiled clips, follow-up refinements, task completion). Implicit
behavior must be interpreted cautiously, since interaction does not equal
relevance \citep{joachims2005accurately}. Such signals could support preference
learning \citep{christiano2017deep}, personalized ranking through contextual
bandits \citep{li2010contextual}, and carefully bounded reinforcement learning
\citep{sutton2018reinforcement}; personalized reranking is one practical first
step. Adaptation is also a memory-management policy: deciding when to run a new
analyzer, consolidate repeated observations, revise schemas, rebuild indexes,
or retire stale memory, while preserving source lineage and deletion semantics. That is the outer loop of \S\ref{sec:loops} executed by the system
itself, reinforcement learning over representation and retrieval policy rather
than over model weights alone, under constraints that keep every change
inspectable and reversible.

\paragraph{Toward native temporal representations}
Our position that the image is the native stored unit of visual information
(\S\ref{sec:tokenization}) is a bet on the present, not a law. The present,
however, is notable: on the evidence of \S\ref{sec:eval}, non-native
solutions (general-purpose components composed by retrieval infrastructure)
currently out-retrieve models pretrained natively for video retrieval at the
early cutoffs where evidence work happens. Representations purpose-built for
video have not yet earned back the flexibility they withdraw. How to encode
temporal information in a learned representation therefore remains, in our
view, among the most valuable open problems in the field. Joint-embedding
predictive architectures learn video representations by predicting in latent
rather than pixel space and now transfer to understanding, prediction, and
planning tasks \citep{assran2025vjepa2}; optical context compression shows a
single vision token can carry roughly an order of magnitude more information
than a text token \citep{wei2025deepseekocr}, a hint of the headroom visual
tokenization retains. What would
change our position is a standard tokenization of visual information across temporal boundaries: a three-dimensional analogue of the text tokenizer that let
language retrieval move into training time. Research in that direction may
find useful precedent in time-series foundation models, which confronted the
same question for continuous signals and answered it with patch-based
tokenizations learned at scale \citep{das2024decoder,ansari2024chronos};
visual streams add two spatial dimensions to that problem, and video
transformers' spatio-temporal tokens are early, mutually incompatible answers
\citep{arnab2021vivit}. If a standard emerges, part of what is orchestration
today will migrate into training time. The VDB format is designed to
survive that migration rather than be displaced by it: a native temporal
representation enters as a new analyzer family, one more layer over the same
sources, while memory, provenance, indexes, and evidence remain the durable
substrate.

\paragraph{Governance}
Persistent collection-wide memory raises access-control, privacy, retention,
and deletion requirements that span artifacts, indexes, cached context,
answers, and evidence streams. Derived representations are not innocuous:
dense embeddings can expose substantial information about source content
\citep{morris2023embeddinginversion}. Deleting a source must therefore have a
defined meaning for every derivative, while removing the source alone does not
guarantee removal of information incorporated into a learned model
\citep{guo2020certifiedremoval}. Video analytics systems also show that privacy
constraints can be enforced at the query layer rather than delegated entirely
to individual analyzers \citep{cangialosi2022privid}. This report treats
governance as a requirement of the category, not a solved problem of the
implementation.

\section{Conclusion}\label{sec:conclusion}

Search over the visual world is not document retrieval with larger files:
its corpus grows while being queried, its unit of meaning is the interval,
its understanding is a portfolio of models, and its answers must remain
playable and attributable. We have given the infrastructure these conditions
demand a precise shape: analyzers define scenes; scenes carry persistent
artifacts; artifacts with provenance constitute visual memory;
capability-declared indexes make memory searchable; search selects bounded
context; and source-linked context becomes evidence that streams. The VDB
format carries this shape in production, and it changes what a search result
\emph{is}: any span of any source, from any timestamp to any timestamp
across a collection, archived or still live, can be named, retrieved,
composed, and watched within seconds, without a media file ever being
touched. When fixed files and fixed segments leave the read path, video
stops behaving like storage and starts behaving like data.

The empirical study locates today’s leverage. Across 9{,}800+ queries, a pipeline of general-purpose
components composed by infrastructure out-retrieved a commercial video-native foundation model at
the strict cutoffs where evidence work happens. The lesson is architectural rather than adversarial:
specialization is something a system can assemble from replaceable parts, not something that must live
inside one indivisible model, and this infrastructure absorbs video-native models as analyzers the day the
balance shifts. One challenge stays open: memory that improves itself through use without weakening
provenance, control, or the right to be forgotten. The question this report closes is what a search over the
visual world returns: evidence one can press play on.

\clearpage
\vspace{4pt}
\bibliographystyle{plainnat}
\bibliography{refs}

\clearpage
\section*{Appendix}
\appendix
\titleformat{\section}{\Large\bfseries}{\thesection.}{0.7em}{}
\section{Exact Dataset-Specific Prompts}\label{app:benchmark-prompts}

The following listings reproduce the complete VideoDB understanding prompts
used in the benchmark, including their required output schemas. Every output
field shown here was stored as text and included in the corresponding semantic
index.

\subsection{YouCook2}
\begin{lstlisting}[style=benchmarkprompt]
You are an expert multimedia-content analyst. For each cooking moment, integrate ordered visual frames, available
narration, on-screen text, and detected ingredients/tools. Output one unified JSON object using the schema below.

===============================================================================
GLOBAL RULES
===============================================================================
1. Multi-Modal Fusion: combine visible preparation with narration and text; state conflicts without inventing facts.
2. Evidence Discipline: report only supported ingredients, quantities, tools, techniques, and state changes.
3. Retrieval Style: queries are imperative recipe steps. Prioritize the exact cooking verb, ingredient,
   tool/container, order of operations, heat treatment, texture/shape change, and resulting food state.
4. Strict JSON Only: return exactly the JSON object. Each description must contain at most two concise sentences.

===============================================================================
FIELD-BY-FIELD EXPECTATIONS
===============================================================================
A. scene_description: unified visual description of the cooking action, ingredients, tools, and result.
B. action: exact preparation action using specific cooking verbs and operand ingredients.
C. recipe_step: Concise imperative instruction matching how a query is written.
D. ingredients: all visible or explicitly narrated ingredients used in this moment, including supported quantities.
E. tools: utensils, cookware, appliances, and containers directly used.
F. food_state: before-to-after condition, including cut shape, mixture consistency, doneness, temperature, or plating.
G. on_screen_text: exact legible ingredient names, quantities, temperatures, timers, or step labels.

===============================================================================
OUTPUT JSON SCHEMA (RETURN EXACTLY THIS STRUCTURE)
===============================================================================
{
  "scene_description": "A cook whisks eggs, flour, black pepper, salt, and cayenne powder together in a metal bowl until combined.",
  "action": "The cook rapidly whisks the eggs and dry seasonings in the bowl to form an even mixture.",
  "recipe_step": "Whisk the eggs, flour, black pepper, salt, and cayenne powder until combined.",
  "ingredients": "eggs, flour, black pepper, salt, cayenne powder",
  "tools": "metal mixing bowl, wire whisk",
  "food_state": "Separate eggs and dry ingredients become a smooth seasoned mixture.",
  "on_screen_text": "none observed"
}

Every text field is required. Use "none observed" when evidence is absent; never return null or an empty string.
\end{lstlisting}

\subsection{MSVD}
\begin{lstlisting}[style=benchmarkprompt]
You are an expert multimedia-content analyst. For each clip, integrate ordered visual frames, available transcript,
on-screen text, and detected objects. Output one unified JSON object using the schema below.

===============================================================================
GLOBAL RULES
===============================================================================
1. Multi-Modal Fusion: use all available evidence together and report conflicts briefly without inventing facts.
2. Evidence Discipline: describe only observable subjects, actions, objects, and locations.
3. Retrieval Style: queries are very short subject-verb-object captions. Prioritize concrete everyday actions,
   animals, cooking, instruments/music, sports, movement, and object manipulation using simple natural wording.
4. Strict JSON Only: return exactly the JSON object. Each description must contain at most two concise sentences.

===============================================================================
FIELD-BY-FIELD EXPECTATIONS
===============================================================================
A. scene_description: unified visual account of subject, action, object, location, activity, and content form.
B. caption: One natural subject-verb-object caption, ideally 3-10 words.
C. action: dominant action using precise present-tense verbs and relevant object interaction.
D. location: concise recognizable environment supported by the frames.
E. character_description: key people or animals with count, role, behavior, interaction, appearance, and colors.
F. object_description: salient manipulated or query-distinctive objects.
G. on_screen_text: exact legible text; never guess obscured text.

===============================================================================
OUTPUT JSON SCHEMA (RETURN EXACTLY THIS STRUCTURE)
===============================================================================
{
  "scene_description": "A brown squirrel sits on a stone ledge outdoors and eats a peanut held between its paws.",
  "caption": "a squirrel is eating a peanut",
  "action": "The squirrel holds the peanut with both front paws and bites through its shell.",
  "location": "Exterior park or garden with a stone ledge and greenery.",
  "character_description": "One small brown squirrel sits upright and focuses on the peanut.",
  "object_description": "A peanut in its shell is held close to the squirrel's mouth.",
  "on_screen_text": "none observed"
}

Every text field is required. Use "none observed" when evidence is absent; never return null or an empty string.
\end{lstlisting}

\subsection{VATEX}
\begin{lstlisting}[style=benchmarkprompt]
You are an expert multimedia-content analyst. For each clip, integrate ordered visual frames, available transcript,
on-screen text, and detected objects. Output one unified JSON object using the schema below.

===============================================================================
GLOBAL RULES
===============================================================================
1. Multi-Modal Fusion: combine all available evidence and state conflicts briefly without inventing facts.
2. Evidence Discipline: include only supported subjects, actions, attributes, topics, and settings.
3. Retrieval Style: captions are comparatively rich and diverse. Capture multi-step actions, interactions,
   how-to activity, clothing/colors, tools, objects, environments, performance, sport, and visible outcomes.
4. Strict JSON Only: return exactly the JSON object. Each description must contain at most two concise sentences.

===============================================================================
FIELD-BY-FIELD EXPECTATIONS
===============================================================================
A. scene_description: rich but concise visual narrative joining subjects, actions, objects, setting, activity/topic,
   and content form.
B. caption: One natural caption, ideally 10-20 words, preserving multiple related actions.
C. action: principal and secondary actions in order, with specific verbs, interactions, and outcome.
D. location: supported environment, spatial layout, lighting, and setting cues.
E. character_description: key people/animals with count, role, behavior, interaction, appearance, clothing, and colors.
F. object_description: salient tools, materials, vehicles, instruments, food, or equipment and their use.
G. on_screen_text: exact legible titles, captions, signs, labels, or interface text.
H. speech_summary: concise spoken topic or instruction; use "none observed" if unavailable.

===============================================================================
OUTPUT JSON SCHEMA (RETURN EXACTLY THIS STRUCTURE)
===============================================================================
{
  "scene_description": "A young man stands at a table and displays shoe polish, water, a soft cloth, and a brush before beginning a shoe-care demonstration.",
  "caption": "a young man shows the supplies needed to polish a shoe",
  "action": "The man lifts each cleaning item toward the camera and arranges the supplies beside the shoe.",
  "location": "Interior instructional setting with a work table and neutral background.",
  "character_description": "One young man faces the camera and presents each item with deliberate hand gestures.",
  "object_description": "A container of shoe polish, water, an old soft cloth, a brush, and a shoe are arranged on the table.",
  "on_screen_text": "none observed",
  "speech_summary": "The presenter identifies the supplies required to polish a shoe."
}

Every text field is required. Use "none observed" when evidence is absent; never return null or an empty string.
\end{lstlisting}

\subsection{MSR-VTT}
\begin{lstlisting}[style=benchmarkprompt]
You are an expert multimedia-content analyst.
For each video clip, integrate evidence from three sources simultaneously:

- Visual frames in logical order
- Full available transcript, including dialogue, sound effects, and spoken text
- On-screen text and detected objects when available

Your task is to output one unified JSON object using the schema at the end.

===============================================================================
GLOBAL RULES
===============================================================================
1. Multi-Modal Fusion
   Analyze every field using visual, transcript, on-screen-text, and object evidence together. If cues conflict,
   state the conflict briefly; do not invent a compromise.

2. Evidence Discipline
   Use only visible or audible evidence. Do not invent identities, titles, intent, locations, or events outside
   the clip. Include a detail only when it improves retrieval.

3. Retrieval Style
   Queries are usually short declarative captions. Prioritize subject + action + object + location and use
   natural query wording. Give special attention to speech/reviews/news, vehicles, music/dance, sports, gameplay,
   animation, cooking/how-to, animals, and everyday human actions.

4. Strict JSON Only
   Return exactly the JSON object and no prose or Markdown outside it. Every description must be concise and contain
   no more than two sentences.

===============================================================================
FIELD-BY-FIELD EXPECTATIONS
===============================================================================
A. scene_description: unified factual visual narrative covering the complete scene, thematic activity, and content
   form such as gameplay, animation, performance, news, tutorial, or movie/TV when evident.
B. caption
   Write one natural declarative caption, ideally 5-15 words, centered on the dominant subject and action.
C. action: dominant visible action with specific verbs, interaction, spatial relationship, pace, and outcome.
D. location: supported interior, exterior, or fantastical setting and distinctive environmental cues.
E. character_description: key people, groups, animals, or animated/game characters with count, appearance, and role.
F. object_description: salient manipulated or query-distinctive objects, vehicles, tools, food, or equipment.
G. on_screen_text
   Perform OCR and return exact legible titles, captions, labels, signs, chart text, or gameplay HUD text in reading
   order. Do not guess obscured text.
H. speech_summary
   Summarize the spoken topic and communicative purpose, including review, interview, news, explanation, or
   demonstration; if no meaningful speech is available, return "none observed".

===============================================================================
OUTPUT JSON SCHEMA (RETURN EXACTLY THIS STRUCTURE)
===============================================================================
{
  "scene_description": "A cook stands behind a kitchen counter beside a bowl of batter and a hot frying pan while demonstrating a pancake recipe.",
  "caption": "a man demonstrates how to cook pancakes",
  "action": "The cook pours batter into the pan, waits for it to set, and flips the pancake with a spatula.",
  "location": "Interior commercial kitchen with stainless-steel appliances and a preparation counter.",
  "character_description": "One adult man in a white chef coat faces the camera and cooks behind the counter.",
  "object_description": "A black frying pan, metal spatula, bowl of batter, and portable stovetop are used in the demonstration.",
  "on_screen_text": "PANCAKE BASICS",
  "speech_summary": "The cook explains when to pour the batter and how to tell when the pancake is ready to flip."
}

Every text field is required. When evidence is absent, return the literal string "none observed"; never return null
or an empty string.
\end{lstlisting}

\paragraph{TwelveLabs}
The Marengo~3.0 indexing interface used in this benchmark does not support
dataset-specific prompts. TwelveLabs therefore has no corresponding prompt
listing; the same provider-wide configuration described in
\S\ref{sec:benchmark-config} was used for every dataset.

\section{Detailed Reranker Study}\label{app:reranker-study}

The reranker study holds each query's 50 semantic candidates fixed and varies
only the reranker and its input mode. Tables~\ref{tab:reranker-msvd}--
\ref{tab:reranker-msrvtt} report the resulting dataset-level recall. The
\emph{No reranking} row preserves the semantic-retrieval order. R@50 is
unchanged within each dataset because every configuration reorders the same
candidate set. The unweighted averages of these values and the corresponding
median component latencies appear in
Table~\ref{tab:reranker-operating-points}.

\begin{table}[H]
\centering
\footnotesize
\setlength{\tabcolsep}{5pt}
\begin{tabular}{@{}llrrrr@{}}
\toprule
\textbf{Model} & \textbf{Input mode} & \textbf{R@1} & \textbf{R@3} & \textbf{R@10} & \textbf{R@50} \\
\midrule
No reranking & -- & 67.34 & 68.63 & 75.69 & 96.39 \\
BGE Gemma & Match Exact & 66.14 & 77.68 & 88.09 & 96.39 \\
BGE Gemma & Concatenate & 70.10 & 80.73 & 89.17 & 96.39 \\
mxbai Base v2 & Match Exact & 66.86 & 71.50 & 82.90 & 96.39 \\
mxbai Base v2 & Concatenate & 69.33 & 76.61 & 86.09 & 96.39 \\
mxbai Large v2 & Match Exact & 67.45 & 70.21 & 79.31 & 96.39 \\
mxbai Large v2 & Concatenate & 68.68 & 72.28 & 81.74 & 96.39 \\
mxbai Edge ColBERT & Match Exact & 64.77 & 76.21 & 87.13 & 96.39 \\
mxbai Edge ColBERT & Concatenate & 66.89 & 77.55 & 86.81 & 96.39 \\
\bottomrule
\end{tabular}
\caption{Dataset-level reranker results on MSVD. Values are percentages.}
\label{tab:reranker-msvd}
\end{table}

\begin{table}[H]
\centering
\footnotesize
\setlength{\tabcolsep}{5pt}
\begin{tabular}{@{}llrrrr@{}}
\toprule
\textbf{Model} & \textbf{Input mode} & \textbf{R@1} & \textbf{R@3} & \textbf{R@10} & \textbf{R@50} \\
\midrule
No reranking & -- & 53.19 & 55.45 & 67.69 & 97.87 \\
BGE Gemma & Match Exact & 62.63 & 79.79 & 93.09 & 97.87 \\
BGE Gemma & Concatenate & 65.96 & 80.98 & 93.48 & 97.87 \\
mxbai Base v2 & Match Exact & 56.78 & 69.55 & 86.70 & 97.87 \\
mxbai Base v2 & Concatenate & 62.50 & 76.06 & 90.16 & 97.87 \\
mxbai Large v2 & Match Exact & 53.72 & 57.85 & 74.47 & 97.87 \\
mxbai Large v2 & Concatenate & 54.92 & 61.04 & 78.72 & 97.87 \\
mxbai Edge ColBERT & Match Exact & 60.90 & 76.60 & 90.29 & 97.87 \\
mxbai Edge ColBERT & Concatenate & 64.36 & 79.39 & 91.36 & 97.87 \\
\bottomrule
\end{tabular}
\caption{Dataset-level reranker results on YouCook2. Values are percentages.}
\label{tab:reranker-youcook2}
\end{table}

\begin{table}[H]
\centering
\footnotesize
\setlength{\tabcolsep}{5pt}
\begin{tabular}{@{}llrrrr@{}}
\toprule
\textbf{Model} & \textbf{Input mode} & \textbf{R@1} & \textbf{R@3} & \textbf{R@10} & \textbf{R@50} \\
\midrule
No reranking & -- & 76.16 & 76.91 & 81.67 & 97.77 \\
BGE Gemma & Match Exact & 78.26 & 87.81 & 94.50 & 97.77 \\
BGE Gemma & Concatenate & 83.46 & 90.28 & 95.24 & 97.77 \\
mxbai Base v2 & Match Exact & 76.67 & 81.80 & 90.21 & 97.77 \\
mxbai Base v2 & Concatenate & 78.93 & 85.79 & 92.67 & 97.77 \\
mxbai Large v2 & Match Exact & 76.98 & 79.57 & 86.43 & 97.77 \\
mxbai Large v2 & Concatenate & 78.19 & 81.50 & 88.45 & 97.77 \\
mxbai Edge ColBERT & Match Exact & 71.94 & 83.86 & 92.47 & 97.77 \\
mxbai Edge ColBERT & Concatenate & 78.12 & 87.00 & 93.42 & 97.77 \\
\bottomrule
\end{tabular}
\caption{Dataset-level reranker results on VATEX. Values are percentages.}
\label{tab:reranker-vatex}
\end{table}

\begin{table}[H]
\centering
\footnotesize
\setlength{\tabcolsep}{5pt}
\begin{tabular}{@{}llrrrr@{}}
\toprule
\textbf{Model} & \textbf{Input mode} & \textbf{R@1} & \textbf{R@3} & \textbf{R@10} & \textbf{R@50} \\
\midrule
No reranking & -- & 63.11 & 64.08 & 68.45 & 92.23 \\
BGE Gemma & Match Exact & 63.59 & 77.67 & 84.47 & 92.23 \\
BGE Gemma & Concatenate & 72.82 & 81.55 & 86.89 & 92.23 \\
mxbai Base v2 & Match Exact & 64.08 & 68.93 & 78.64 & 92.23 \\
mxbai Base v2 & Concatenate & 66.99 & 77.18 & 83.01 & 92.23 \\
mxbai Large v2 & Match Exact & 64.08 & 67.48 & 75.24 & 92.23 \\
mxbai Large v2 & Concatenate & 66.99 & 72.82 & 80.58 & 92.23 \\
mxbai Edge ColBERT & Match Exact & 62.62 & 73.30 & 84.95 & 92.23 \\
mxbai Edge ColBERT & Concatenate & 66.02 & 77.18 & 84.95 & 92.23 \\
\bottomrule
\end{tabular}
\caption{Dataset-level reranker results on MSR-VTT. Values are percentages.}
\label{tab:reranker-msrvtt}
\end{table}

\section{Operational Measurement Details}\label{app:operational}

The latency measurements in this appendix originated from the same client
region. Requests from a different region may add approximately 100--200~ms.

\subsection{Direct structured and aggregate access}

We measured direct \emph{Query} and \emph{Aggregate} over five hours of
indexed video. Both contracts operated collection-wide and applied the same
eligibility filter. The following examples show representative request forms.
The reported statistics contain 200 calls: 100 \emph{Query} calls and 100
\emph{Aggregate} calls.

\begin{lstlisting}[style=benchmarkprompt,language=Python]
eligibility_filter = [
    {
        "field": "outputs.attribute",
        "op": "exists",
        "value": True,
    }
]

query_results = collection.query(
    index_name="scene_query_aggregate_latency_v1",
    filter=eligibility_filter,
    limit=100,
    return_fields=["outputs.attribute"],
)

aggregate_results = collection.aggregate(
    index_name="scene_query_aggregate_latency_v1",
    filter=eligibility_filter,
    group_by="outputs.attribute",
    metric="count",
    limit=100,
)
\end{lstlisting}

\begin{table}[H]
\centering
\small
\begin{tabular}{@{}lrr@{}}
\toprule
\textbf{Contract} & \textbf{Measured calls} & \textbf{Median latency (s)} \\
\midrule
\emph{Query} & 100 & 0.443 \\
\emph{Aggregate} & 100 & 0.380 \\
\bottomrule
\end{tabular}
\caption{Median latency for direct structured and aggregate access over five
hours of indexed video.}
\label{tab:direct-operational}
\end{table}

\subsection{Evidence-stream generation}

Table~\ref{tab:stream-operational} reports median generation latency for
playable evidence streams assembled from one source and from five sources.

\begin{table}[H]
\centering
\small
\begin{tabular}{@{}rcc@{}}
\toprule
\textbf{Duration} & \textbf{Single-source median} & \textbf{Five-source median} \\
\midrule
5 s   & 441 ms & 469 ms \\
10 s  & 433 ms & 433 ms \\
30 s  & 443 ms & 495 ms \\
1 min & 434 ms & 447 ms \\
2 min & 521 ms & 595 ms \\
5 min & 588 ms & 600 ms \\
10 min & 728 ms & 994 ms \\
\bottomrule
\end{tabular}
\caption{Median generation latency for playable evidence streams assembled
from one source and from five sources.}
\label{tab:stream-operational}
\end{table}

Requested stream duration increased by 120$\times$, from five seconds to ten
minutes, while median client-observed generation latency increased by only
1.65$\times$ for single-source streams (441 to 728~ms) and 2.12$\times$ for
five-source streams (469 to 994~ms). Generation latency therefore grew much
more slowly than requested stream duration. Across every measured duration and
source count, the client-observed median remained below one second, giving a
temporal search result a sub-second median path to playable evidence.

\end{document}